\documentclass[aip,amsmath,amssymb,preprint]{revtex4-1}

\usepackage{graphicx}
\usepackage{dcolumn}
\usepackage{bm}
\usepackage[utf8]{inputenc}
\usepackage[T1]{fontenc}
\usepackage{etoolbox}
\usepackage[separate-uncertainty = true, multi-part-units = repeat, per-mode=symbol]{siunitx}
\usepackage{xcolor}
\usepackage{lipsum}
\usepackage{hyperref}
\usepackage[normalem]{ulem}
\draft 

\begin{document}

\title{Hyperfine and Zeeman interactions in ultracold collisions of molecular hydrogen with atomic lithium}

\author{Hubert~J\'{o}\'{z}wiak}
\email{hubert.jozwiak@doktorant.umk.pl}
 \affiliation{%
 Institute of Physics, Faculty of Physics, Astronomy and Informatics, Nicolaus Copernicus University in Toru\'{n}, Grudziadzka 5, 87-100 Toru\'{n}, Poland}%
 \email{hubert.jozwiak@doktorant.umk.pl}
 
 \author{Timur V. Tscherbul}%
 \affiliation{
Department of Physics, University of Nevada, Reno, NV 89557, USA}%

\author{Piotr~Wcis{\l}o$^{1}$}%

\date{\today}

\begin{abstract}
We present {a} rigorous quantum scattering study of the effects of hyperfine and Zeeman interactions on cold Li -- H$_{2}$ collisions in the presence of an external magnetic field using a recent \textit{ab initio} potential energy surface. We {find that the low-field-seeking states of H$_{2}$ predominantly undergo elastic collisions: the ratio of elastic-to-inelastic {cross-sections} exceeds 100 for collision energies below {100~mK}. Furthermore, we demonstrate that most inelastic collisions conserve the space-fixed projection of the nuclear spin. We show that the {anisotropic hyperfine} interaction between the nuclear spin of H$_{2}$ and the electron spin of Li {can have a 
 significant} effect on inelastic scattering in the ultracold regime, as it mediates two processes: the electron spin relaxation in lithium, and the nuclear spin -- electron spin exchange. {G}iven the predominance of elastic collisions and the propensity of inelastic collisions to retain H$_{2}$ in {its} low-field-seeking states, {our results} open up the possibility of sympathetic cooling of molecular hydrogen by atomic lithium, paving the way for future exploration of ultracold collisions and high-precision spectroscopy of H$_{2}$ molecules.} 
\end{abstract}

\maketitle


\section{\label{sec:introduction}Introduction}
Cold collisions {and chemical reactions} involving molecular hydrogen have been the subject of {much theoretical and experimental} interest {due to their significance in} astrochemistry {and cold} controlled chemistry.\cite{Krems_2008, Balakrishnan_2016} {In particular, t}he $\mathrm{F+H_{2}}\rightarrow \mathrm{HF+H}$ reaction, despite having a high energetic barrier of approximately 800~K, occurs quite efficiently in cold environments (10--100~K) and is the only known source of HF in the interstellar medium.~\cite{Tizniti_2014} Experiments on the Penning ionization of H$_{2}$ upon collisions with metastable ($^{3}\mathrm{S}$) helium revealed sub-K shape resonances,~\cite{Henson_2012} isotopic effects,~\cite{Lavert_Ofir_2014} and a significant role of molecular rotation~\cite{Shagam_2015} and anisotropy of the molecule-atom interaction~\cite{Klein_2016} in cold reaction dynamics. Cold collisions of vibrationally excited isotopologues of molecular hydrogen (HD and D$_{2}$) with D$_{2}$,~\cite{Perreault_2017, Perreault_2018} H$_{2}$,~\cite{Perreault_2018} and He~\cite{Zhou_2021_Science,Zhou_2021}~revealed {interesting} stereodynamic effects,~\cite{Jambrina_2019} interference patterns and shape resonances, which enable {quantum interference-based coherent control of the collision outcome}.\cite{Devolder_2020, Devolder_2021}

Previous theoretical studies of cold collisions involving molecular hydrogen and its isotopologues~\cite{Balakrishnan_1997, Balakrishnan_1998, Mack_2006, Quemener_2008, Quemener_2009, Balakrishnan_2011, Santos_2011, Croft_2018, Croft_2019, Jambrina_2019, Morita_2020, Jambrina_2022} neglected the effects of hyperfine interactions and Zeeman shifts on collisions with H$_{2}$, which could be substantial at ultralow temperatures. For instance, the hyperfine splitting of the $\nu=0, N=1$ state in \textit{ortho}-H$_{2}$ is approximately~600~kHz~\cite{Ramsey_1952,Jozwiak_2020, Puchlaski_2020} (or $k_{B}\,20\,\mu$K). The {hyperfine} structure of this state in an external magnetic field is quite complex, {comprising} nine Zeeman states.~\cite{Ramsey_1952} {However, ultracold c}ollision dynamics involving these states, and the mechanisms driving particular transitions (e.g. nuclear spin relaxation) remain unexplored.

An additional motivation to study {the role of hyperfine and Zeeman interactions in} cold collisions of H$_{2}$ molecules is related to {high-precision} spectroscopy of molecular hydrogen. {Accurate} determination of energy intervals between rovibrational states in hydrogen (with relative accuracy reaching the sub-ppb level~\cite{Fast2020,Fast_2021}) allows for performing stringent tests of quantum electrodynamics~\cite{Komasa2019,Puchalski2019,Zaborowski_2020} and for putting constraints on the strength of hypothetical {interactions} beyond the Standard Model.~\cite{Ubachs2016} To overcome Doppler broadening {and enhance the precision of the determined transition frequencies}, experimental groups employ saturation techniques,~\cite{Diouf2019,Hua_2020} molecular beams~\cite{Fast2020,Fast_2021} or cooling of the gas sample~\cite{Kassi_2022, Cozijn_2022, Liu_2022} (down to 57~K). {Further improvement in high-precision} spectroscopy {would be possible if} molecular hydrogen {could be} cooled and trapped. Recently, we proposed a scheme for implementing a magic wavelength for fundamental transition $\nu=0, N=0 \rightarrow \nu=1, N=2$ in \textit{para}-H$_{2}$~\cite{Jozwiak_2022} ({which is not magnetically trappable}) {that has a potential to enable much higher accuracy}. In contrast, magnetic {trapping} could be used to increase the precision spectroscopy of the fundamental transition ($\nu=0, N=1 \rightarrow \nu=1, N=1$) in \textit{ortho}-H$_{2}$. Both the optical dipole and magnetic traps could reach depths {of the order of} 1~mK, for the laser power density of 1~MW/mm$^{2}$ and magnetic field strength {of} 0.4~T, respectively {(as recently demonstrated in Ref.\cite{Singh_2023} it is possible to use strong and focused laser beams to achieve trap depths of approximately 3.6~K for H$_{2}$)}. Further progress in high-precision spectroscopy {is contingent upon the ability to cool} H$_{2}$ far below 1~mK.

One such {possibility} is sympathetic cooling, which relies on immersing a molecular system in a gas of coolant atoms,  preferably of a similar mass.\cite{Lara_2006,Lara_2007,Tscherbul_2011b, Morita2017} Under such conditions, elastic collisions {result in cooling by} transferring momentum between molecules and the coolant atoms. In contrast, inelastic collisions can cause transitions to high-field-seeking (untrappable) states, {which} release {the} internal energy {stored in these states}, {and} lead to heating and trap loss.\cite{Tscherbul_2011b, Morita2017} For efficient sympathetic cooling, the ratio of the cross-section for elastic to inelastic collisions ($\gamma = \sigma_{\rm{el}}/\sigma_{\rm{inel}}$) should be larger than 100.\cite{Carr_2009}

A relatively small mass and the fact that it can be cooled down to the $\mu$K regime using laser cooling techniques make atomic lithium {an attractive} candidate for sympathetic cooling of H$_{2}$. Duarte~\textit{et~al.}~\cite{Duarte_2011} have demonstrated {a} magnetooptical trap (MOT) {for $^{6}$Li atoms} operating on a narrow $2S_{1/2}\rightarrow 3P_{3/2}$ transition at 323~nm, {achieving} temperatures as low as 59~$\mu$K.  {However, t}he feasibility of sympathetic cooling of H$_{2}$ {by} collisions with Li remains to be determined. For instance, there is a significant mismatch in the Zeeman splittings of the two species, which could lead to losses once {the} lithium MOT and a hypothetical magnetic trap of H$_{2}$ are overlapped.  {Furthermore, it is unclear whether} the ratio, $\gamma$, of the cross-section for elastic to inelastic collisions of H$_{2}$ with Li is high enough to support the cooling process. In order to accurately calculate $\gamma$ it is necessary to take into account the effects of hyperfine interactions and the presence of an external magnetic field.

In this paper, we present the first rigorous theoretical study of {the role of} hyperfine and Zeeman interactions effects {in cold} atom -- H$_{2}$ collisions. We investigate cold collisions of \textit{ortho}-H$_{2}$ ($\nu=0, N=1$) {molecules} with $^{6}$Li {atoms} using coupled-channel quantum scattering calculations based on a {highly accurate} \textit{ab initio} potential energy surface (PES).~\cite{Makrides2019} In the field-free case, we find that the three hyperfine states {of H$_{2}$} are collisionally stable, i.e.~the ratio of cross-sections for elastic ($F$-conserving) to inelastic ({$F$-changing}) collisions exceeds 100, with the exception of a narrow range in the vicinity of {a} $g$-wave shape resonance located at $E\approx k_{B}\times1.2\,\rm{K}$. We find that the presence of {the} magnetic dipolar interaction between the nuclear magnetic moment of H$_{2}$ and the electron spin magnetic moment of Li manifests itself in the ultracold regime, where it drastically increases the inelastic $F=0 \rightarrow F'=1$ scattering (otherwise suppressed by the selection rules {for} transitions driven by the H$_{2}$-Li interaction potential). {We also} perform quantum scattering calculations in an external magnetic field, and we analyze relaxation from the three low-field-seeking Zeeman states in H$_{2}$ {upon collisions with $^{6}$Li atoms in the trappable ($S=1/2, M_{S}=1/2$) state}. We find that the inelastic {relaxation} is dominated by collisions which conserve the space-fixed projection of the nuclear spin of H$_{2}$. The {magnetic dipolar} interaction drives the relaxation of the electronic spin of Li, which has profound consequences on the possibility of sympathetic cooling of H$_{2}$ by lithium. 

The article is organised as follows. In Sec.~\ref{sec:Theory} we outline the quantum theory of collisions between $^{1}\Sigma$ molecules {with two magnetic nuclei (such as \textit{ortho}-H$_{2}$)} and $^{2}S$ atoms in an external magnetic field, which includes the intramolecular hyperfine interactions in the $^{1}\Sigma$ molecules, as well as the spin-dependent interaction between the molecule and the $^{2}S$ atom. Next, we apply this theory to {cold Li -- H$_{2}$} collisions {in Sec.~\ref{sec:results:field_free}, where} we present and discuss the results for the field-free case. {Then, in Sec.~\ref{sec:results:magnetic_field} we provide the state-to-state cross sections for transitions between {the} Zeeman sublevels {of \textit{ortho}-H$_{2}$} in an external magnetic field. We discuss the kinetic energy and magnetic field dependence of the cross-sections, as well as the implications of the calculated elastic-to-inelastic scattering ratio on the possibility of sympathetic cooling of H$_{2}$ by $^{6}$Li.} Sec.~\ref{sec:conclusions} we conclude {by summarizing the main results of this} work. Atomic units are used throughout the article unless stated otherwise.

\section{Theory}
\label{sec:Theory}
In this section, we {present} the quantum theory of {collisions between} a $^{1}\Sigma$ molecule and a $^{2}S$ atom in a presence of an external magnetic field. The theory is based on the seminal works of Krems and Dalgarno~\cite{Krems_2004} and Volpi and Bohn,~\cite{Volpi_2002} who first considered diatom -- atom collisions in a magnetic field. It is also an extension of the recent work of Hermsmeier~\textit{et~al}.~\cite{Hermsmeier_2023} (who {studied} nuclear spin relaxation in {cold He-$^{13}$C$^{16}$O} collisions), to the case of collisions {of open-shell atoms, such as $^{6}$Li, and molecules with two magnetic nuclei, such as \textit{ortho}-H$_{2}$.}

We use space-fixed Jacobi coordinates to describe the scattering system: the  separation vector $\mathbf{R}$ from the atom to the center of mass of the H$_{2}$ molecule, the internuclear vector $\mathbf{r}$, and {the angle $\theta$ between} $\mathbf{R}$ and $\mathbf{r}$. There are six {angular momenta} in the H$_{2}$($^{1}\Sigma_{g}^{+}$)-Li($^{2}S$) system: the rotational angular momentum of the nuclei in H$_{2}$ ($\hat{\mathbf{N}}$) the nuclear spin angular momenta of the protons, $\hat{\mathbf{I}}_{1}$ and $\hat{\mathbf{I}}_{2}$ ($I_{i} = |\hat{\mathbf{I}}_{i}| = 1/2$, $i=1,2$), the total electron spin of the lithium atom, $\hat{\mathbf{S}}$ ($S = |\hat{\mathbf{S}}| = 1/2$), the nuclear spin angular momentum of lithium, $\hat{\mathbf{I}}_{\rm{Li}}$ (${I}_{\rm{Li}} = |\hat{\mathbf{{I}}}_{\rm{Li}}| = 1$ for $^{6}$Li and $3/2$ for $^{7}$Li), and the angular momentum operator describing the {orbital} motion of the {collision partners}, $\hat{\boldsymbol{l}}$. For reasons {clarified below}, we neglect the hyperfine structure of lithium and we exclude $\hat{\mathbf{I}}_{\rm{Li}}$ from the analysis. We couple the two nuclear spins of the protons to form the total nuclear spin of H$_{2}$, $\hat{\mathbf{I}}_{\rm{H_{2}}} = \hat{\mathbf{I}}_{1} + \hat{\mathbf{I}}_{2}$. We recall that molecular hydrogen exists in two {spin} isomeric forms: \textit{para}-H$_{2}$, with ${I}_{\rm{H_{2}}} = |\hat{\mathbf{I}}_{\rm{H_{2}}}|=0$  and \textit{ortho}-H$_{2}$ with ${I}_{\rm{H_{2}}} = |\hat{\mathbf{I}}_{\rm{H_{2}}}|=1$. {Because} the total wavefunction of {H$_{2}$ must be} antisymmetric with respect to the permutation of the protons, \textit{para}-H$_{2}$ exhibits rotational structure with only even values of $N=|\hat{\mathbf{N}}|$, while the rotational structure of \textit{ortho}-H$_{2}$ involves only odd $N$ values. {Since $I_{\rm{H_{2}}}=0$ for \textit{para}-H$_{2}$, this spin isomer does not have the hyperfine structure and nuclear Zeeman shifts, and interacts with magnetic field only through its weak rotational magnetic moment ({for $N=0$ \textit{para}-H$_{2}$, the rotational magnetic moment is strictly zero}). In contrast, \textit{ortho}-H$_{2}$ does have a non-zero nuclear spin, yielding stronger Zeeman shifts that make it amenable to magnetic trapping. Thus, we focus on  \textit{ortho}-H$_{2}$ {in what follows}.}

The Hamiltonian for the {atom-molecule collision} system is
\begin{align}
    \begin{split}
    \label{eq:Hamiltonian}
        \hat{H} =  - \frac{1}{2\mu R} \frac{\partial^{2}}{\partial R^{2}} R + \frac{\hat{\boldsymbol{l}}^{2}}{2\mu R^{2}} + \hat{V}(\mathbf{R},\mathbf{r}) + \hat{V}_{\rm{SD}}(\mathbf{R},\mathbf{r},\hat{\mathbf{I}},\hat{\mathbf{S}}) + \hat{H}_{\rm{as}},
    \end{split}
\end{align}
where ${\mu = m_{\rm{at}}m_{\rm{mol}}/(m_{\rm{at}}+m_{\rm{mol}})}$ is the reduced mass of the {collision partners} {(we use $m_{\rm{at}} = 6.015121$ and $m_{\rm{mol}} = 2.01565$ atomic mass units)\cite{Brown_2003}},  $\hat{V}(\mathbf{R},\mathbf{r})$ is the {atom}-molecule potential energy surface, $\hat{V}_{\rm{SD}}(\mathbf{R},\mathbf{r},\hat{\mathbf{I}},\hat{\mathbf{S}})$ denotes the spin-dependent (SD) {Hamiltonian} ({note that the subscript in $\hat{\mathbf{I}}_{\rm{H_{2}}}$ is dropped for simplicity, and the nuclear spin of H$_{2}$ is denoted simply as $\hat{\mathbf{I}}$})
\begin{align}
    \begin{split}
    \label{eq:SDpart}
        \hat{V}_{\rm{SD}}(\mathbf{R},\mathbf{r},\hat{\mathbf{I}},\hat{\mathbf{S}}) = \sum_{i=1,2}A^{i}_{F}(\mathbf{R},\mathbf{r})\hat{\mathbf{I}}_{i}\cdot\hat{\mathbf{S}}
        + \sum_{i=1,2}\sum_{\alpha,\beta}c^{i}_{\alpha\beta}(\mathbf{R},\mathbf{r}) \hat{\mathbf{I}}_{i_{\alpha}}\hat{\mathbf{S}}_{\beta} ,
    \end{split}
\end{align}
where the sum over $i$ involves the two protons in H$_{2}$, and the $\alpha$ and $\beta$ run over {Cartesian components of the spin operators in a} molecule-fixed coordinate {frame}. The first term corresponds to the Fermi contact interaction between the nuclear spin angular momenta, $\hat{\mathbf{I}}_{i}$ of the $i$-th proton in H$_{2}$ and the spin angular momentum, $\hat{\mathbf{S}}$, of the {valence} electron in lithium, with $A^{i}_{F}(\mathbf{R},\mathbf{r})$ being the coupling coefficient for the Fermi {contact} interaction. {Due to the similarity in the interaction potentials and reduced masses, the magnitude of the Fermi contact interaction in H$_{2}$ -- Li can be estimated from the previous work on $^{3}$He -- Li\cite{Tscherbul_2009} and $^{3}$He -- K\cite{Tscherbul_2011a} collisions. At the zero-energy turning point of the Li -- H$_{2}$ potential ($R\approx 8.75\,a_{0}$), the Fermi contact interaction constant for $^{3}$He -- Li\cite{Tscherbul_2009} and $^{3}$He -- K\cite{Tscherbul_2011a} is on the order of $10^{-4}$~cm$^{-1}$.} Since this interaction vanishes rapidly with increasing {$R$},~\cite{Tscherbul_2011a} its influence on {the} low-temperature {Li -- H$_{2}$} scattering is expected to be negligible. We thus exclude the Fermi contact interaction from our analysis. The second term in Eq.~\eqref{eq:SDpart} is the {intermolecular anisotropic hyperfine} interaction, the strength of which is determined by the coupling tensor, $c^{i}_{\alpha\beta}(\mathbf{R},\mathbf{r})$. Since calculating the full dependence of the coupling tensor on $R$, {$\theta$} and $r$ is beyond the scope of this work, we use an approximate formula which is appropriate for describing the long-range part of the {anisotropic hyperfine} interaction. We assume that the total nuclear spin magnetic moment of H$_{2}$, $\hat{\boldsymbol{\mu}}_{\rm{H_{2}}} = g_{H}\mu_{N} \hat{\mathbf{I}}$, and the electron spin magnetic moment of Li,  $\hat{\boldsymbol{\mu}}_{\rm{Li}} = g_{S}\mu_{B} \hat{\mathbf{S}}$ are point dipoles located at the centers of mass of H$_{2}$ and Li, respectively. The magnetic dipole interaction between the two magnetic moments is given as\cite{Tscherbul_2011a}
\begin{align}
    \begin{split}
    \label{eq:SDpart-dipole}
        \hat{V}_{\rm{SD}}(\mathbf{R},\hat{\mathbf{I}},\hat{\mathbf{S}}) = -g_{S}\mu_{B}g_{\mathrm{H}}\mu_{N} \sqrt{\frac{24\pi}{5}}\frac{\alpha^{2}}{R^{3}}  \sum_{q=-2}^{2}(-1)^{q}Y_{2,-q}(\hat{\mathbf{R}}) \Bigl[\hat{\mathbf{S}}\otimes\hat{\mathbf{I}}\Bigr]^{2}_{q} ,
    \end{split}
\end{align}
where $g_{S}$ and $g_{H}$ are the electron and proton $g$-factors, respectively, $\mu_{B}$ and $\mu_{N}$ denote the Bohr and nuclear magnetons, and $\alpha$ is the fine-structure constant. $Y_{2q}(\hat{\mathbf{R}})$ is a spherical harmonic of rank 2, which depends on the orientation of the scattering system, and $\Bigl[\hat{\mathbf{S}}\otimes\hat{\mathbf{I}}\Bigr]^{2}_{q}$ is a tensorial product of $\hat{\mathbf{S}}$ and $\hat{\mathbf{I}}$. {We note that the general {expression} for the {anisotropic hyperfine} interactions, Eq.~\eqref{eq:SDpart}, is used in studies of hyperfine and Zeeman effects in three-atom molecules which involve nuclear and electronic spins, such as HCO,\cite{Bowater_1973} NH$_{2}$,\cite{Cook_1977} and Na$_{3}$.\cite{Coudert_2002,Hauser_2015} {These interactions also play a crucial role in electron spin decoherence of alkali-metal atoms trapped in solid \textit{para}-H$_{2}$ matrices.}\cite{Upadhyay2019} A form similar to Eq.~\eqref{eq:SDpart-dipole} is used to describe the {long-range} magnetic {dipolar} interaction between {the} electron {spins} of $^{2}\Sigma$ molecules and $^{2}S$ atoms,\cite{Tscherbul_2011b, Morita2017} two $^{2}\Sigma$ molecules\cite{Krems_2004} and two $^{3}\Sigma$ molecules.\cite{van_der_Avoird_1987,Krems_2004,Janssen_2011, Suleimanov_2012}}

The asymptotic Hamiltonian {$\hat{H}_{\rm{as}}$ in Eq.~\eqref{eq:Hamiltonian} is given by}
\begin{equation}
\label{eq:asymptHamiltonian}
    \hat{H}_{\rm{as}} = \hat{H}_{\rm{H_{2}}} + \hat{H}_{\rm{Li}},
\end{equation}
where $\hat{H}_{\rm{H_{2}}}$ and $\hat{H}_{\rm{Li}}$ correspond to the Hamiltonians of the isolated molecule and atom, respectively. The effective Hamiltonian  for the H$_{2}$ molecule in the ground electronic ($^{1}\Sigma_{g}^{+}$) state is
\begin{equation}
\label{eq:asympt-H2}
    \hat{H}_{\rm{H_{2}}} = \hat{H}_{\rm{rot}} +
\hat{H}_{\rm{HF}} + \hat{H}_{\rm{Zeeman}},
\end{equation}
and involves the rotational, intramolecular hyperfine (HF), and Zeeman terms:
\begin{equation}
     \hat{H}_{\rm{rot}} = {B_{v}}\hat{\mathbf{N}}^{2} - D_{v}\hat{\mathbf{N}}^{4},
\end{equation}
\begin{equation}
     \hat{H}_{\rm{HF}} =-c_{\rm{nsr}}\hat{\mathbf{N}}\cdot \hat{\mathbf{I}} + g_{\rm{H}}^{2}\mu_{\rm{N}}^{2}\Bigl(\frac{\mu_{0}}{4\pi}\Bigr) \Biggl(\frac{\hat{\mathbf{I}}_{1}\cdot\hat{\mathbf{I}}_{2}}{r^{3}}-\frac{3(\hat{\mathbf{I}}_{1}\cdot\mathbf{r})(\hat{\mathbf{I}}_{2}\cdot\mathbf{r})}{r^{5}}\Biggr),
\end{equation}
\begin{equation}
\label{eq:H2Zeeman}
 \hat{H}_{\rm{Zeeman}} = -g_{\rm{r}} \mu_{\rm{N}} \hat{N}_{Z}{B}_{Z} - g_{\rm{H}} \mu_{\rm{N}}{I}_{{Z}}{B}_{Z}(1-\sigma). 
\end{equation}
Here, {$B_{v}$ and $D_{v}$ are the effective rotational and centrifugal distortion constants in vibrational state $v$}. The intramolecular hyperfine Hamiltonian {describes} the two dominant hyperfine interactions in H$_{2}$ -- the nuclear spin-rotation interaction, and the {dipolar} interaction {between the nuclear spins}. The respective hyperfine coupling constants, {$c_{\rm{nsr}}$ and $c_{\rm{dip}}$, quantify} the strength of these two interactions. The two terms in the Zeeman Hamiltonian correspond to the contribution of the interaction of the rotational magnetic moment and the nuclear magnetic moment with the external magnetic field, with $g_{\rm{r}}$, and $\sigma$ being the rotational nuclear $g$-factor and the anisotropic part of the nuclear shielding tensor, respectively. We assume that the external magnetic field is aligned along the space-fixed Z-axis. The diamagnetic interaction parameterized by molecular susceptibility, i.e. the interaction of the magnetic field with an induced molecular magnetic moment, contributes significantly only in intense magnetic fields ($B > 1$~T)\cite{Hermsmeier_2023} and thus we {neglect} it in the following analysis.

{Because our interest here is in transitions between the hyperfine states of H$_{2}$}, we {also} neglect the internal hyperfine structure of the lithium atom. Thus, the effective Hamiltonian for the isolated lithium atom, $\hat{H}_{\rm{Li}}$, involves only the Zeeman term
\begin{equation}
\label{eq:asympt-Li}
    \hat{H}_{\rm{Li}} = -g_{S} \mu_{B} \hat{S}_{{Z}} B_{Z} ,
\end{equation}
where $g_{S}$ is the electron spin $g$-factor.
 
{The total wavefunction of the system is expanded in a complete set of uncoupled basis {states} in {a} space-fixed frame of reference\cite{Krems_2004,Tscherbul_2007}
\begin{align}
\begin{split}
\label{eq:total_wv}
    |\Psi\rangle = \frac{1}{R}\sum_{N} \sum_{M_{N}=-N}^{N}&\sum_{M_{I}=-1}^{1} \sum_{M_{S}=-1/2}^{1/2} \sum_{l}\sum_{M_{l}=-l}^{l} F_{N M_{N} M_{I} M_{S} l M_{l}} (R) |N M_{N} \rangle |I M_{I} \rangle  |S M_{S} \rangle |lM_{l}\rangle,
\end{split}
\end{align}
$M_{N}$, $M_{I}$, $M_{S}$, and $M_{l}$ are the projections of $\hat{\mathbf{N}}$, $\hat{\mathbf{I}}$, $\hat{\mathbf{S}}$,  and $\hat{\boldsymbol{l}}$ on the space-fixed $Z$-axis, respectively.} {The expansion \eqref{eq:total_wv} is appropriate for weakly anisotropic atom -- molecule interaction potentials, such as {the} Li -- H$_{2}$ potential used in this work.}

Substitution of the total wavefunction from Eq.~\eqref{eq:total_wv} to the Schr\"{o}dinger equation, ${\hat{H}|\Psi\rangle = E|\Psi\rangle}$ leads to a set of coupled channel (CC) equations for the expansion coefficients, {$F_{N M_{N} M_{I} M_{S} l M_{l}} (R)$}
\begin{widetext}
\begin{align}
    \begin{split}
    \label{eq:CC}
        \Bigl[\frac{d^{2}}{dR^{2}}+ & 2\mu E - \frac{l(l+1)}{R^{2}}\Bigr]F_{N M_{N} M_{I} M_{S} l M_{l}} (R) = \\
         =2\mu&\sum_{N',M_{N}',M_{I}',M_{S}',l',M_{l}'}   F_{N' M_{N}' M_{I}' M_{S}' l' M_{l}'} (R) \\
         \times & \langle N M_{N}|\langle I M_{I}| \langle S M_{S}| \langle l M_{l}| \hat{V}(\mathbf{R},\mathbf{r}) + \hat{V}_{\rm{SD}}(\mathbf{R},\mathbf{r},\hat{\mathbf{I}},\hat{\mathbf{S}}) + \hat{H}_{\rm{as}}|N' M_{N}'\rangle |I M_{I}'\rangle S M_{S}'\rangle | l' M_{l}'\rangle 
    \end{split}
\end{align}
\end{widetext}
{where $E$ is the total energy, and $\mu$ is defined in Eq.~\eqref{eq:Hamiltonian}}. The evaluation of the matrix elements {on the right-hand side} is {described} in Appendix~A. Note that the {CC} equations are block-diagonal with respect to $M = M_{N} + M_{I_{\rm{H_{2}}}} + M_{S} +  M_{l}$, the projection of the total angular momentum, $\hat{\mathbf{J}}$, on the space-fixed Z-axis. This is a consequence of the fact that in the presence of an external magnetic field, $M$, contrary to ${J}$, is conserved.\cite{Krems_2004,Tscherbul_2007} {This allows us to solve the {CC} equations for each value of $M$ separately.}

We solve the CC equations numerically (for computational details see Sec.~\ref{subsec:computational_details}), and transform the asymptotic solution to the eigenstate basis {of the asymptotic Hamiltonian,~\eqref{eq:asymptHamiltonian}} {for H$_{2}$ in {a magnetic} field} 
\begin{equation}
\label{eq:mixing}
    |(NI)\gamma_{\rm{H_{2}}}\rangle = \sum_{M_{N}=-N}^{N}\sum_{M_{I}=-I}^{I} A^{\gamma_{\rm{H_{2}}}}_{M_{I},M_{N}}(B) |N M_{N} \rangle |I M_{I} \rangle,
\end{equation}
{where $\gamma_{\rm{H_{2}}}$ denotes the eigenvalue {of the H$_{2}$ Hamiltonian~\eqref{eq:asympt-H2}}.} In principle, the asymptotic Hamiltonian involves a term that couples different rotational states of H$_{2}$, {but} this coupling {is extremally small, as shown in Appendix~\ref{sec:AppendixA}, so} we treat $N$ as a good quantum number. We also note that the asymptotic Hamiltonian {of the Li atom}, introduced in Eq.~\eqref{eq:asympt-Li} is diagonal {{in the basis of} $|S M_{S}\rangle$ states, {thus $|S M_{S}\rangle$ is an approximate eigenvector for an isolated lithium atom {with $M_{S} = \pm 1/2$ labeling the atomic Zeeman levels}}. Next, we match the result to the linear combinations of the Riccati-Bessel and Neumann functions {to} obtain the scattering S-matrix.\cite{Johnson_1973} {T}he state-to-state cross sections {are calculated} from the S-matrix elements at a given collision energy, $E_{\rm{kin}}$, {by} {summing contributions from all $M$-blocks\cite{Krems_2004}
\begin{align}
\begin{split}
         \sigma_{\gamma_{\mathrm{H_{2}}} M_{S} \rightarrow \gamma_{\rm{H_{2}}}'M_{S}'} (E_{\rm{kin}}) = \frac{\pi}{k_{\gamma_{\rm{H_{2}}}M_{S}} ^{2}} \sum_{M}\sum_{l M_{l}} \sum_{l\ M_{l'}} \Bigl|\delta_{l,l'}\delta_{M_{l},M_{l}'}\delta_{\gamma_{\rm{H_{2}}},\gamma_{\rm{H_{2}}}'}\delta_{M_{S},M_{S}'} - S^{M}_{\gamma_{\rm{H_{2}}}M_{S} l M_{l}, \gamma_{\rm{H_{2}}}'M_{S}' l' M_{l}'} \Bigr|^{2} ,
\end{split}
\end{align}
where $k_{\gamma_{{\rm{H_{2}}}\,M_{S}}}=\sqrt{2\mu(E-E_{\gamma_{\rm{H_{2}}}} - E_{M_{S}})}$ is the collision wavevector.} 

Since we are interested in the {collisional} relaxation of the nuclear spin {states of} molecular hydrogen, we define a state-to-state cross-section, which is {summed} over the final {Zeeman} states of the lithium atom
\begin{equation}
\label{eq:simga_av}
    \sigma_{\gamma_{\rm{H_{2}}} \rightarrow \gamma_{\rm{H_{2}}}'} (E_{\rm{kin}}) = \sum_{M_{S}'}\sigma_{\gamma_{\mathrm{H_{2}} {M_{S}=1/2}} \rightarrow \gamma_{\mathrm{H_{2}}}'M_{S}'}  (E_{\rm{kin}}),
\end{equation}
and the related rate coefficient
\begin{align}
\begin{split}
\label{eq:k_av}
    k_{\gamma_{\rm{H_{2}}} \rightarrow \gamma_{\rm{H_{2}}}'}(T) 
 =\sqrt{\frac{8}{\pi \mu k_{B}^{3}T^{3}}} \int_{0}^{\infty} \sigma_{\gamma_{\rm{H_{2}}} \rightarrow \gamma_{\rm{H_{2}}}'} (E_{\rm{kin}}) E_{\rm{kin}}e^{-E_{\rm{kin}}/k_{B}T} \mathrm{d}E_{\rm{kin}} .
\end{split}
\end{align}
{In this work, we assume that the lithium atom is initially in the trappable $M_{S} = 1/2$ state, {so we can drop the } $M_{S}$ symbol on the left-hand side of Eqs.~\eqref{eq:simga_av} and~\eqref{eq:k_av}.}

When considering collisions {in the absence of an} external magnetic field, we {expand the total wavefunction as follows
\begin{align}
\begin{split}
\label{eq:total_wv_coupled}
    |\Psi\rangle = \frac{1}{R}\sum_{N} \sum_{F=|N-1|}^{N+1}&\sum_{M_{S}=-1/2}^{1/2} \sum_{l}\sum_{M_{l}=-|l|}^{l} F_{N F M_{F} M_{S} l M_{l}} (R) |(NI) F M_{F} \rangle |S M_{S} \rangle |lM_{l}\rangle,
\end{split}
\end{align}
i.e. we use the coupled basis vectors to represent the states of H$_{2}$
\begin{equation}
    |(NI)F M_{F}\rangle = (-1)^{-N+1-M_{F}} \sum_{M_{N}=|-N|}^{N}\sum_{M_{I}=-1}^{1}  \sqrt{2F+1} 
    \begin{pmatrix}
        N & 1 & F\\
        M_{N} & M_{I} & -M_{F}
    \end{pmatrix}
    |N M_{N}\rangle |I M_{I}\rangle.
\end{equation}
Here, {
    $\tiny\begin{pmatrix}
        . & . & .\\
        . & . & .
    \end{pmatrix}$ are the 3-j symbols,}\cite{Zare_1988} $F = |\hat{\mathbf{F}}|$ is the quantum number associated with the total angular momentum of H$_{2}$, $\hat{\mathbf{F}}$, which is the result of coupling of $\hat{\mathbf{N}}$ to $\hat{\mathbf{I}}$}. This representation is convenient {because $F$ is conserved} in the field-free case, see Sec.~\ref{sec:results:field_free}. {The rest of the procedure follows the same steps as detailed {above}, and hence is not repeated here. {The m}atrix elements of the PES, {the} spin-dependent interaction, and {the} asymptotic Hamiltonian in the {coupled} basis ({see} Eq.~\eqref{eq:total_wv_coupled}) are provided in Appendix~\ref{sec:appendixB}.}

\subsection{Computational details}
\label{subsec:computational_details}
\begin{table}[!ht]
    \centering
    \begin{tabular}{c|c|c|}
         Constant       & Value & Source   \\ \hline
         {$B_{v}$}        & {59.322~cm$^{-1}$} & \cite{Huber_1979}  \\
         $D_{v}$        & {$4.575\times10^{-2}$~cm$^{-1}$}  & \cite{Huber_1979} \\
         $c_{\rm{nsr}}$ & $(3.81\pm0.01)\times10^{-6}$~cm$^{-1}$  & \cite{Jozwiak_2020}   \\
         $c_{\rm{dip}}$ &  $(9.614\pm0.005)\times10^{-6}$~cm$^{-1}$ &  \cite{Jozwiak_2020} \\
         $g_{\rm{r}}$   &  0.8825 & \cite{Pachucki_2011} \\
         $g_{\rm{H}}$   &  5.5856946983 & \cite{CODATA} \\
         $\sigma$       & $1.76$~ppm  & \cite{Sundholm_1996}  \\
    \end{tabular}
    \caption{Spectroscopic parameters of H$_{2}$ used in the calculations {reported in the present work}.}
    \label{tab:mol_parameters}
\end{table}

The spectroscopic constants used to parameterize the Hamiltonian of H$_{2}$~(Eq.~\eqref{eq:asympt-H2}) and Li~(Eq.~\eqref{eq:asympt-Li}) are {listed} in Table~\ref{tab:mol_parameters}. We use the \textit{ab initio} H$_{2}$-Li PES reported by Makrides~\textit{et~al}.~\cite{Makrides2019} This PES was recently used in calculations of elastic, inelastic, and glancing-angle rate coefficients for collisions of ultracold Li atoms with room-temperature H$_{2}$ {molecules} in the context of the calibration of a cold-atom vacuum standard.~\cite{Booth_2019,Makrides2019,Shen_2023,Klos_2023}  For the purpose of solving the CC equations, we expand the PES in  Legendre polynomials~(Eq.~\eqref{eq:PESexpansion}). Since H$_{2}$ is a homonuclear molecule, the expansion index takes only even values. We truncate the expansion in Eq.~\eqref{eq:PESexpansion} at $\lambda_{\rm{max}} = 4$. {The dependence of the expansion coefficients on the H$_{2}$ stretching coordinate, $r$, is averaged out by the integration over rovibrational wave functions of the isolated H$_{2}$ molecule in the ground vibrational state, see Eq.~\eqref{eq:rovibaverage} for details.}

We solve the CC equations using {a} log-derivative propagator\cite{Johnson_1973,Manolopoulos_1986} on a radial grid from $R_{\rm{min}} = 3.0\,a_{0}$ to $R_{\rm{max}} = 200~a_{0}$ (for collisions with $E_{\mathrm{kin}} < 10^{-2}$\,cm$^{-1}$ we increase $R_{\rm{max}}$ to $500~a_{0}$) with a constant step {size} of $0.05\,a_{0}$. {While this {integration} range is notably smaller than that used in Ref.~\cite{Makrides2019}, where the authors extended $R_{\rm{max}}$ to $5000~a_{0}$, {we validated that it is sufficient} to ensure a subpercent convergence of the state-to-state cross sections.} We cover the range of kinetic energies from $10^{-9}$~cm$^{-1}$ to 50~cm$^{-1}$. {Due to the weak anisotropy of the Li -- H$_{2}$ PES,} it is sufficient to keep only {the two lowest} two rotational levels {of \textit{ortho}-H$_{2}$} ($N=1$ and $N=3$) in the basis to obtain a subpercent convergence of the cross sections. The number of partial waves, $l_{\rm{max}}$, included in {our} calculations depends on the collision energy and varies from $6$ up to $55$. {To verify our calculations we compared the field-free cross-sections with the previous results\cite{Makrides2019} and found excellent agreement.} {Finally, we note that the exact value of the rotational constant has no significant influence on the cross-sections. For instance, switching between $B_{v=0}$ and $B_{e}$ modifies elastic cross-section for scattering of H$_{2}$ in the $F=0$ hyperfine state by 0.03\%, and the inelastic cross-sections by less than 0.005\%.}

\section{Results: field-free H$_{2}$-Li collisions}
\label{sec:results:field_free}

\begin{figure}[!ht]
\centering
\includegraphics[width=0.5\linewidth]{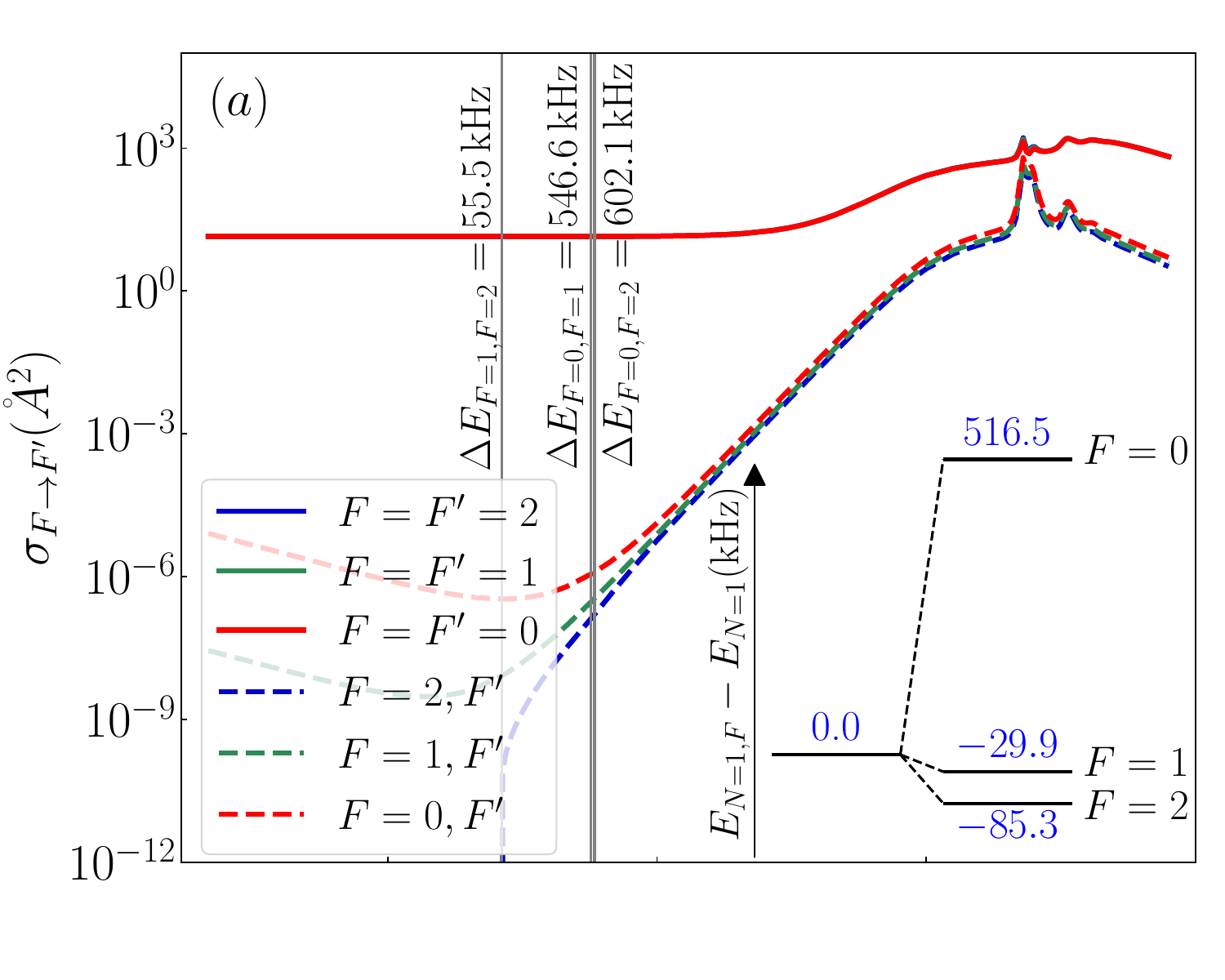}\hspace{-0.12cm}
\includegraphics[width=0.5\linewidth]{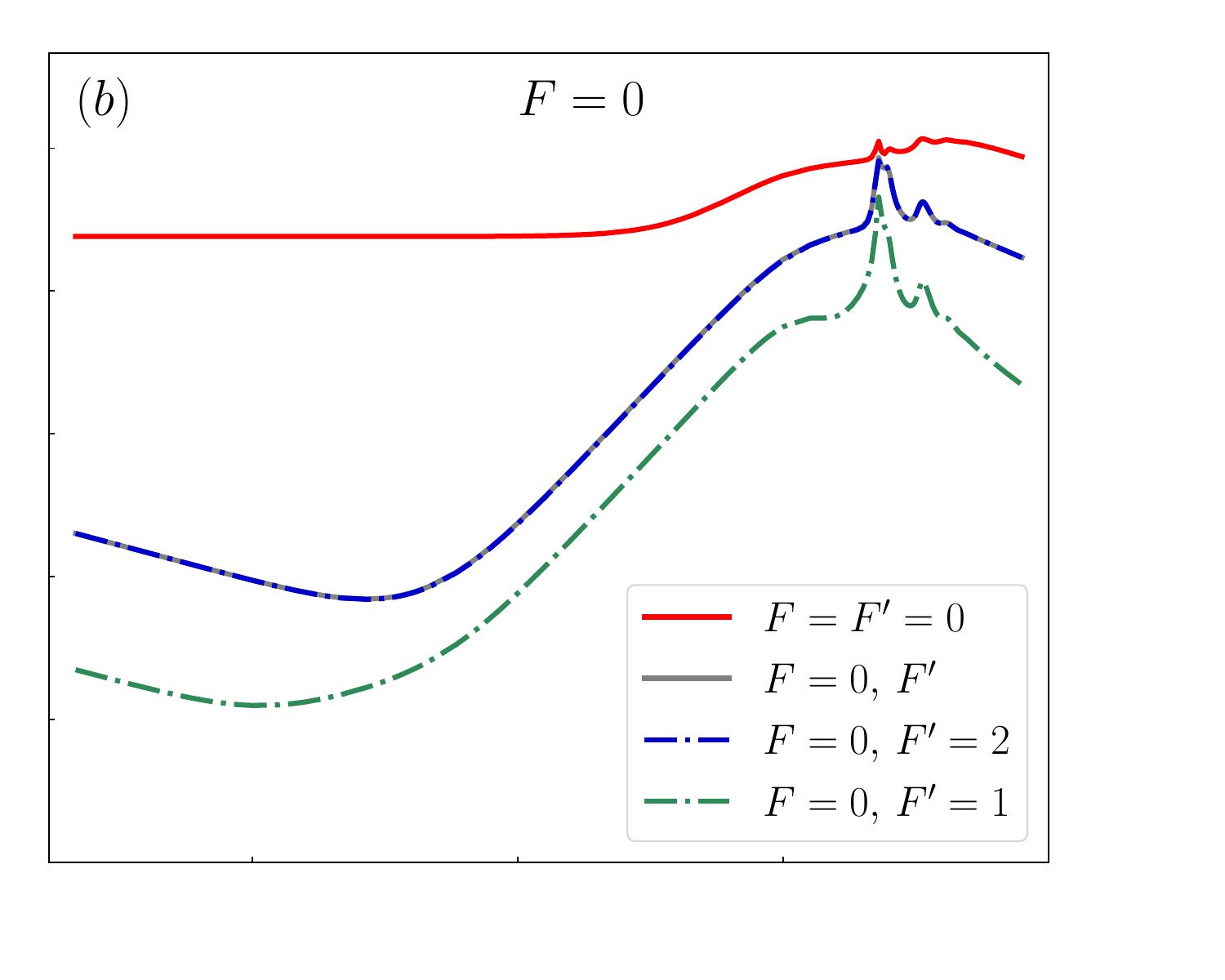}\\\vspace{-0.48cm}
\includegraphics[width=0.5\linewidth]{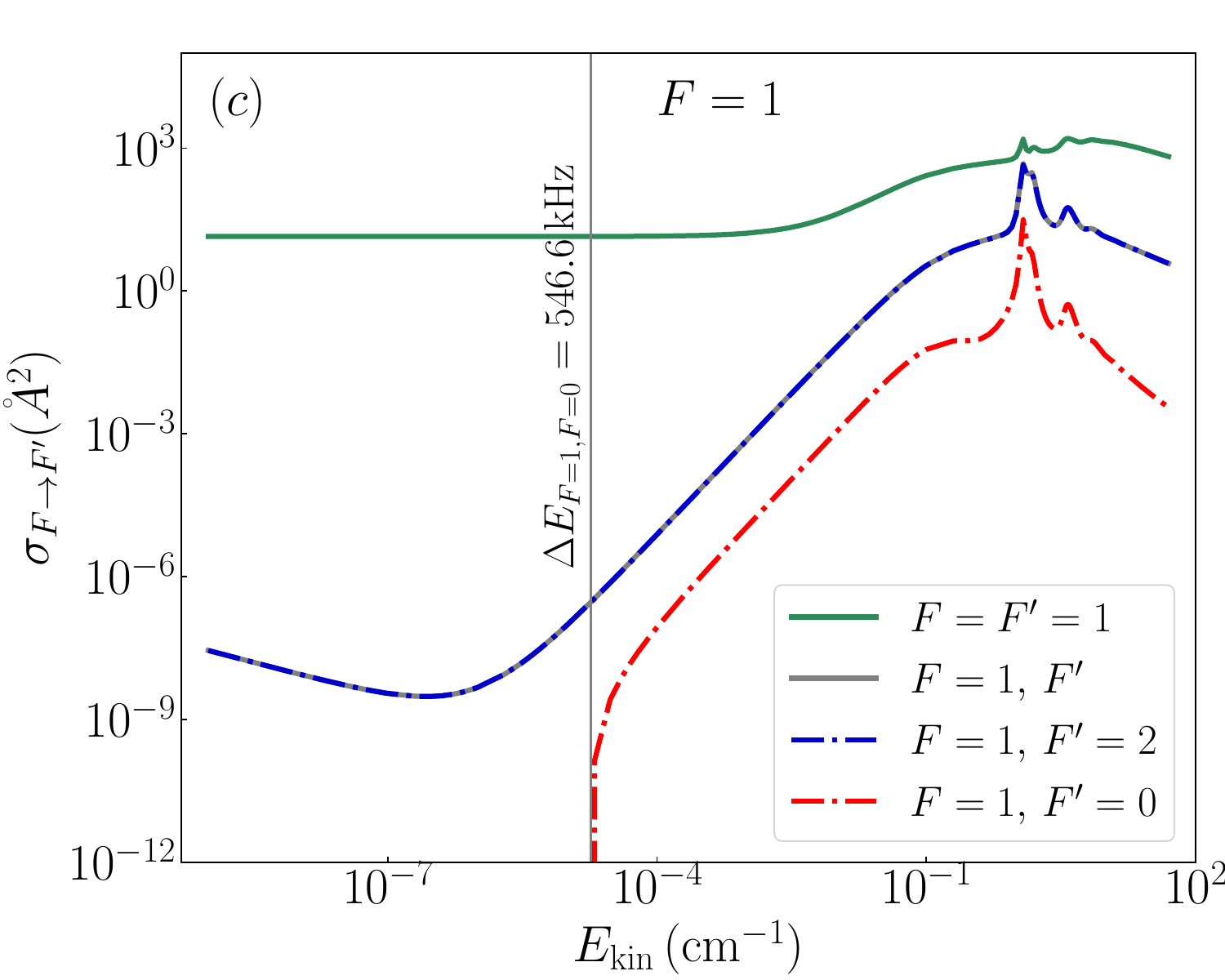}\hspace{-0.12cm}
\includegraphics[width=0.5\linewidth]{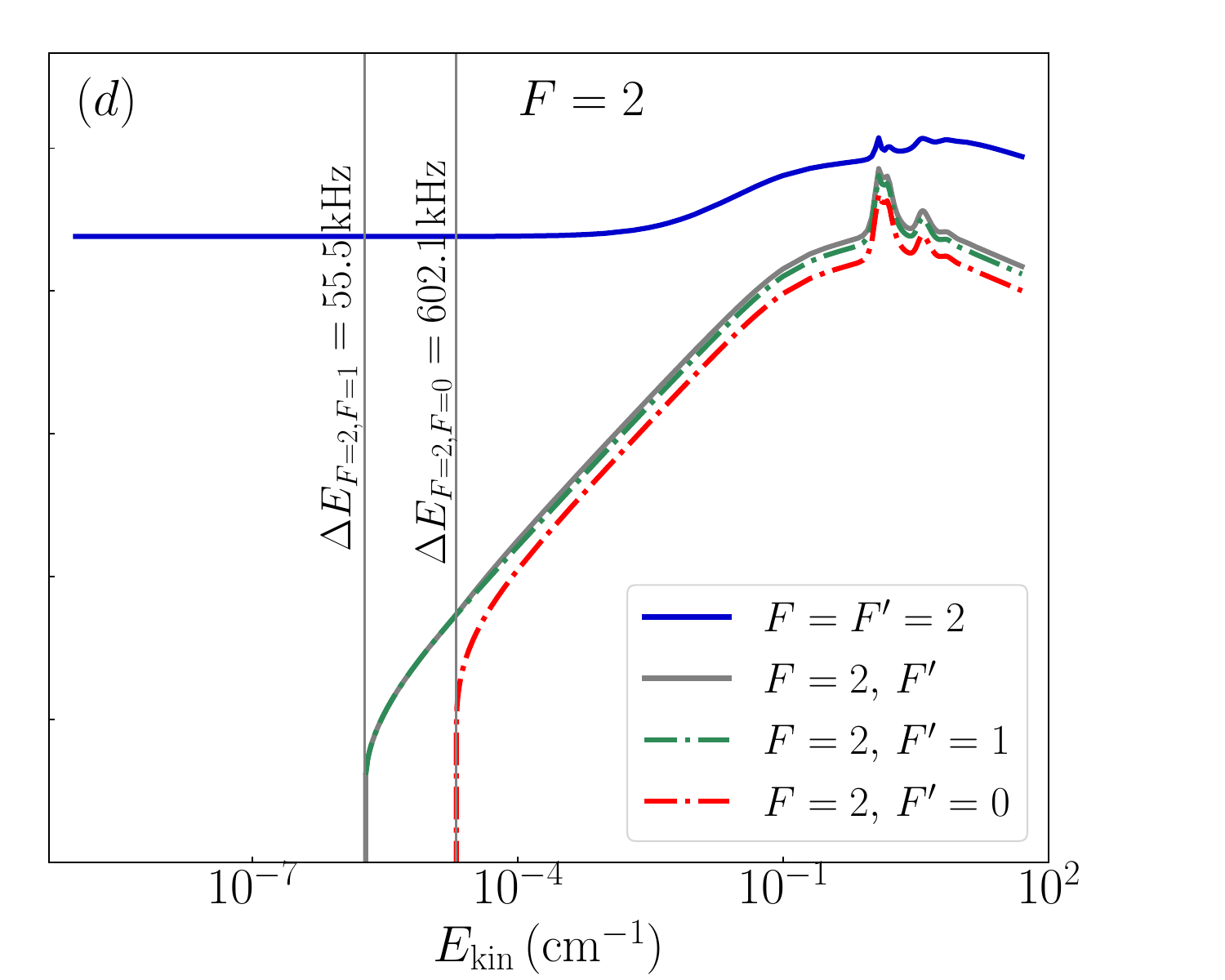}
\caption{Hyperfine-resolved state-to-state cross sections for collisions of H$_{2}$ with $^{6}$Li. Panel ($a$) presents elastic ($F$-conserving) cross sections (solid lines), and the total inelastic state-to-state cross sections (dashed lines) for a given $F$ initial state. The gray vertical lines correspond to {the} energy differences between the three hyperfine states and are shown to {illustrate} the mechanism of channel opening. The energy diagram in the bottom right corner of panel {(a)} {shows} the hyperfine {structure} of the $\nu=0, N=1$ level {of} H$_{2}$. {In panels (b)-(d), we show the cross-sections separately for each initial value of $F$ {along with} the decomposition of the total inelastic cross-section into the different final state, $F^{\prime}$, contributions.}}
\label{fig:FieldFree}
\end{figure}
In the absence of an external magnetic field, the $\nu=0, N=1$  state in H$_{2}$ is split into three hyperfine levels, spread over a range of approximately 600~kHz,\cite{Jozwiak_2020} {as shown in} {the inset {of} Fig.~\ref{fig:FieldFree}~(a)}. {The three states are labeled by $F$, the total angular momentum of H$_{2}$}. We present the hyperfine-resolved state-to-state cross sections of H$_{2}$ ($\nu=0,N=1$) colliding with $^{6}$Li in Fig.~\ref{fig:FieldFree}. The elastic ($F$-conserving) cross sections are shown {in Fig.~1($a$)} as blue ($F=2$), green ($F=1$), and red ($F=0$) lines, respectively. The {cross sections} are almost identical, apart from kinetic energies close to the {feature} located at $E_{\rm{kin}} = 1.2$~cm$^{-1}$, where the largest difference between the cross sections approach $20\%$. {T}his structure was attributed to the $g$-wave ($l=4$) shape resonance in Ref.~\cite{Makrides2019} The inelastic ($F$-changing) cross sections are {typically orders of magnitude smaller than the elastic cross sections. However, near $E_{\rm{kin}}=1.2$~cm$^{-1}$ this difference narrows to a factor of 2.5}. The three panels ($(b), (c)$, and $(d)$) provide additional information about the inelastic processes which affect each $F$-labelled state. Inelastic scattering from the $F=0$ state, which has the largest energy, is dominated by the $F=0\rightarrow F'=2$ deexcitation ({the grey solid and blue dashed-dotted lines in Fig.~1(b)} are almost overlapped). {This process is driven by the $\lambda = 2$ {anisotropic} term of the Li -- H$_{2}$ PES (see Eq.~\eqref{eq:coupled_ME_pes}).} The $F'=1$ state is not directly coupled {to the initial state} by the PES (there is no $\lambda=1$ term in the PES expansion {since H$_{2}$ is a homonuclear molecule}), and the cross-sections for the $F=0\rightarrow F'=1$ deexcitation are at least {one} order of magnitude smaller than {those for the ${F=0 \rightarrow F'=2}$ transition}. The nuclear spin -- electron spin interaction introduces a weak coupling between the $F=0$ and $F=1$ levels, which influences the cross-sections only for kinetic energies smaller than 10$^{-6}$~cm$^{-1}$ {as discussed below}.  Similarly, inelastic scattering from the $F=1$ state is dominated by the $F=1 \rightarrow F'=2$ deexcitation, driven by the $\lambda = 2$ term in the PES expansion. When kinetic energy surpasses the $\Delta E_{F=0,F=1} = E_{F=0} - E_{F=1}$ threshold (at 546.5~kHz), excitation to the $F=0$ state becomes energetically accessible. As mentioned above, since the $F=1$ and $F=0$ {channels} are not directly coupled by the PES, this contribution to the total inelastic cross-section is significantly (two orders of magnitude) weaker than the $F=1 \rightarrow F'=2$ deexcitation. For the $F=2$ {initial} state, we note that until the kinetic energy surpasses the first threshold ($E_{F=1,F=2} = E_{F=1} - E_{F=2}$ at $54.6$~kHz) the scattering is purely $F$-conserving. Both the $F=2 \rightarrow F'=1$ and $F=2 \rightarrow F'=0$ excitations are driven by the $\lambda = 2$ term (with a weak contribution from the spin-dependent interaction).

\begin{figure}
    \centering
    \includegraphics[width=0.5\linewidth]{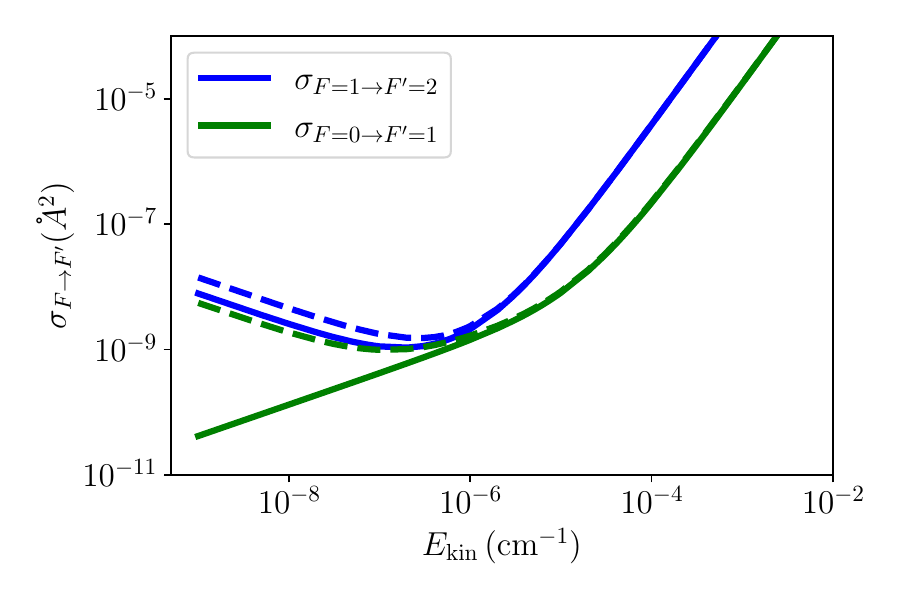}
    \caption{Influence of the nuclear spin--electron spin interaction on the hyperfine-structure resolved state-to-state cross sections for collisions of H$_{2}$ with $^{6}$Li. The dashed and solid lines present the cross-sections calculated including and neglecting the spin-dependent interaction, respectively.}
    \label{fig:MDinteraction}
\end{figure}
Overall, the {magnetic dipole-dipole} interaction, {Eq.~\eqref{eq:SDpart-dipole},} has a negligible influence on the state-to-state cross sections for collisions of H$_{2}$ with $^{6}$Li in the absence of an external magnetic field. The only significant effect that we observe is an enhancement of the $\Delta F = 1$ deexcitation in the ultracold regime (below $10^{-6}$~cm$^{-1}$) -- the dashed and solid lines in Fig.~\ref{fig:MDinteraction} {re}present the values of $\sigma_{F \rightarrow F'=F+1}$ calculated with and without the spin-dependent H$_{2}$-Li interaction, respectively. Interestingly, the inclusion of this weak interaction is necessary to obtain the Wigner threshold behavior\cite{Wigner_1948} of the $\sigma_{F=0 \rightarrow F'=1}$ cross-section ($\sigma \sim E_{\rm{kin}}^{-1/2}$) {at $E_{\rm{kin}} \approx 10^{-6}$~cm$^{-1}$}. We observe a slight alternation of the $\Delta F = -1$ \textit{excitation} cross-sections near the thresholds ($\Delta E_{F=0,F=1} = 546.5$~kHz and $E_{F=1,F=2} = 54.6$~kHz), which {is not shown} in Fig.~\ref{fig:MDinteraction}. In the remaining {field-free} cases (larger relative kinetic energies and other scattering processes) the nuclear spin--electron spin interaction has a negligible influence on the state-to-state cross-sections.

\section{Results: H$_{2}$-Li collisions in an external magnetic field}
\label{sec:results:magnetic_field}

Before proceeding to {discuss} the results of scattering calculations, we briefly {consider} the energy structure of the H$_{2}$ molecule in an external {magnetic} field. {Figure~\ref{fig:fig3}($a$) shows} the Zeeman energy levels {in} the $\nu=0, N = 1$ {rovibrational manifold of \textit{ortho}-H$_{2}$ obtained by diagonalization of the Hamiltonian  in Eq.~\eqref{eq:asympt-H2} with {the} spectroscopic parameters {of H$_{2}$} gathered in Table~\ref{tab:mol_parameters}.} At large field strengths, the nine levels are grouped into sets of three states, which share the same projection of the nuclear spin, $M_{I}$. This reflects the relative strength of the nuclear Zeeman term with respect to the rotational Zeeman term -- for $N=1$, the first term in Eq.~\eqref{eq:H2Zeeman} is approximately six times smaller than the second term. {Within each group of states, the order of states (starting from states with the largest energy) is $M_{N} = -1, 0$, and $1$. At high fields, the top three {low-field-seeking Zeeman} states, which are amenable to magnetic trapping, correspond to $M_{I} = -1$.}

\begin{figure}[!ht]
    \centering
    \includegraphics[width=0.48\linewidth]{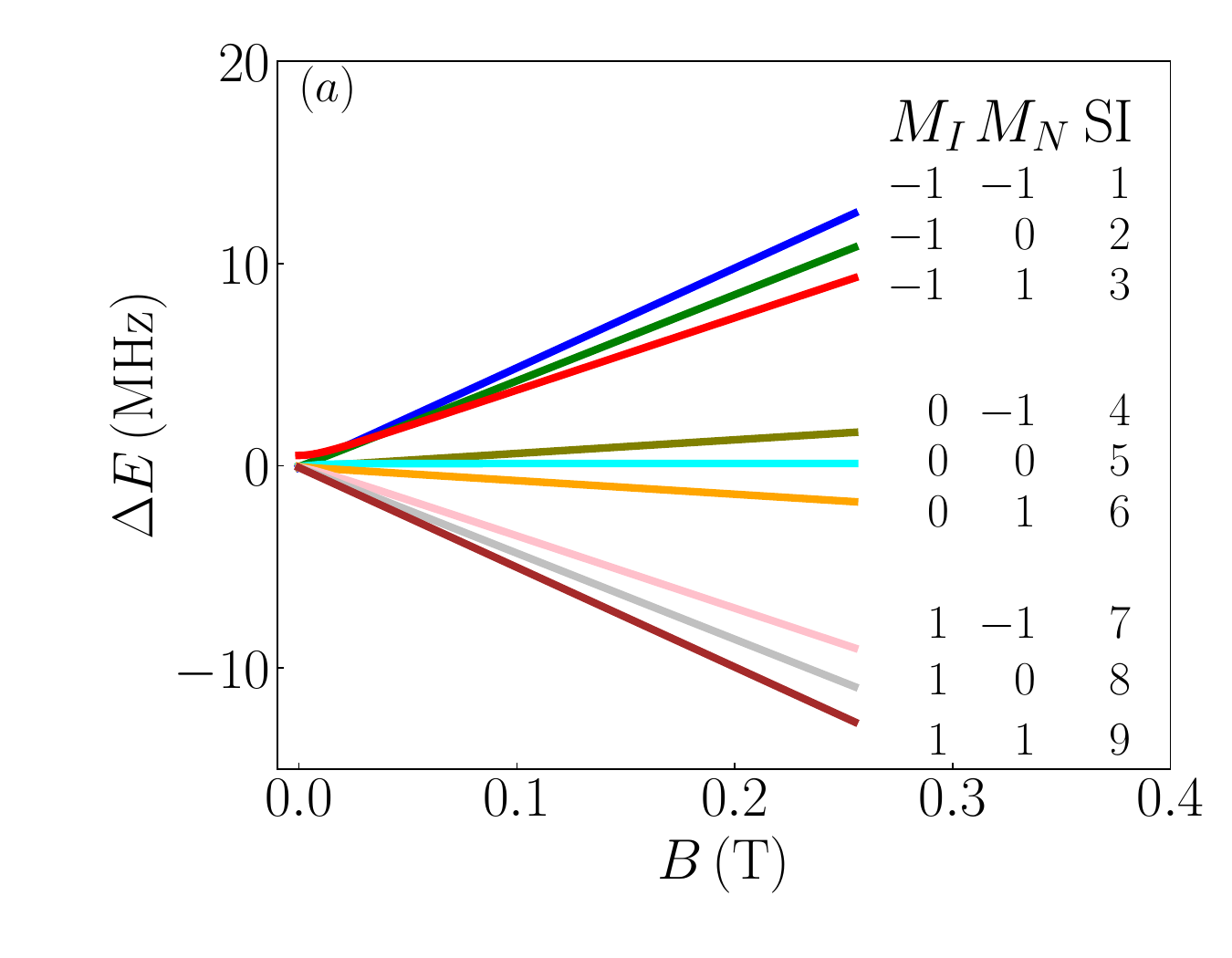}
    \includegraphics[width=0.48\linewidth]{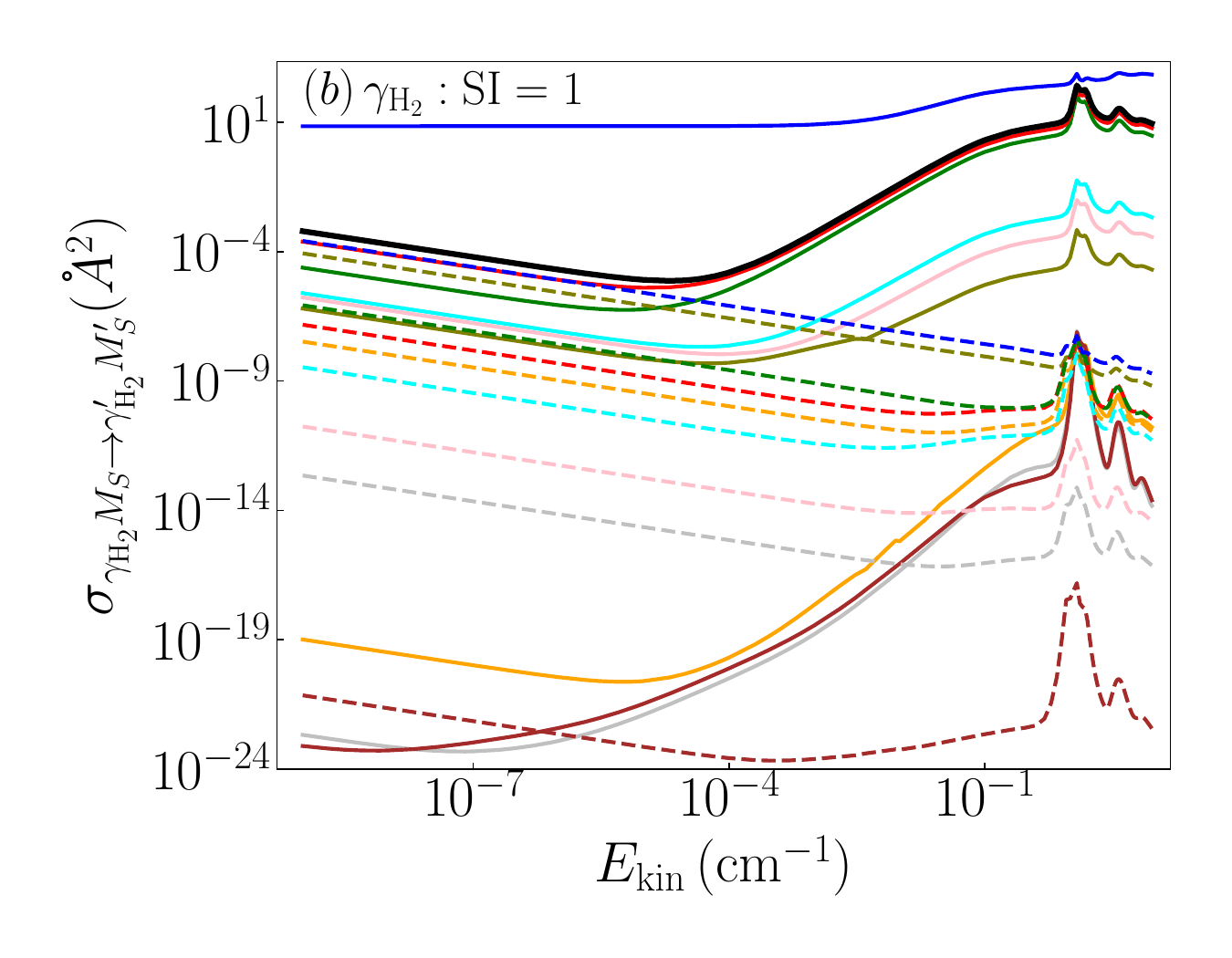}\hspace{-0.4cm} \\
    \includegraphics[width=0.48\linewidth]{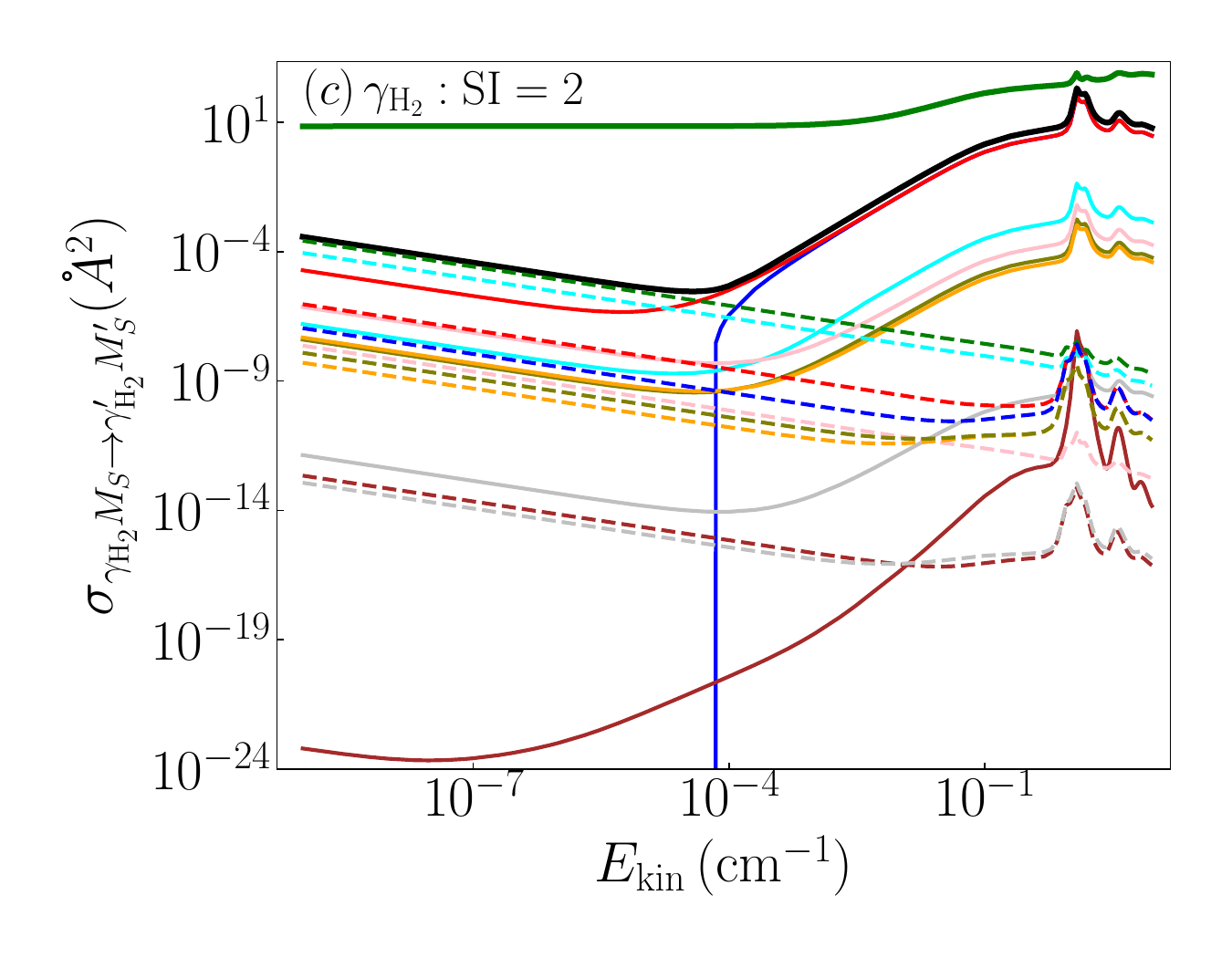}\hspace{-0.12cm}
    \includegraphics[width=0.48\linewidth]{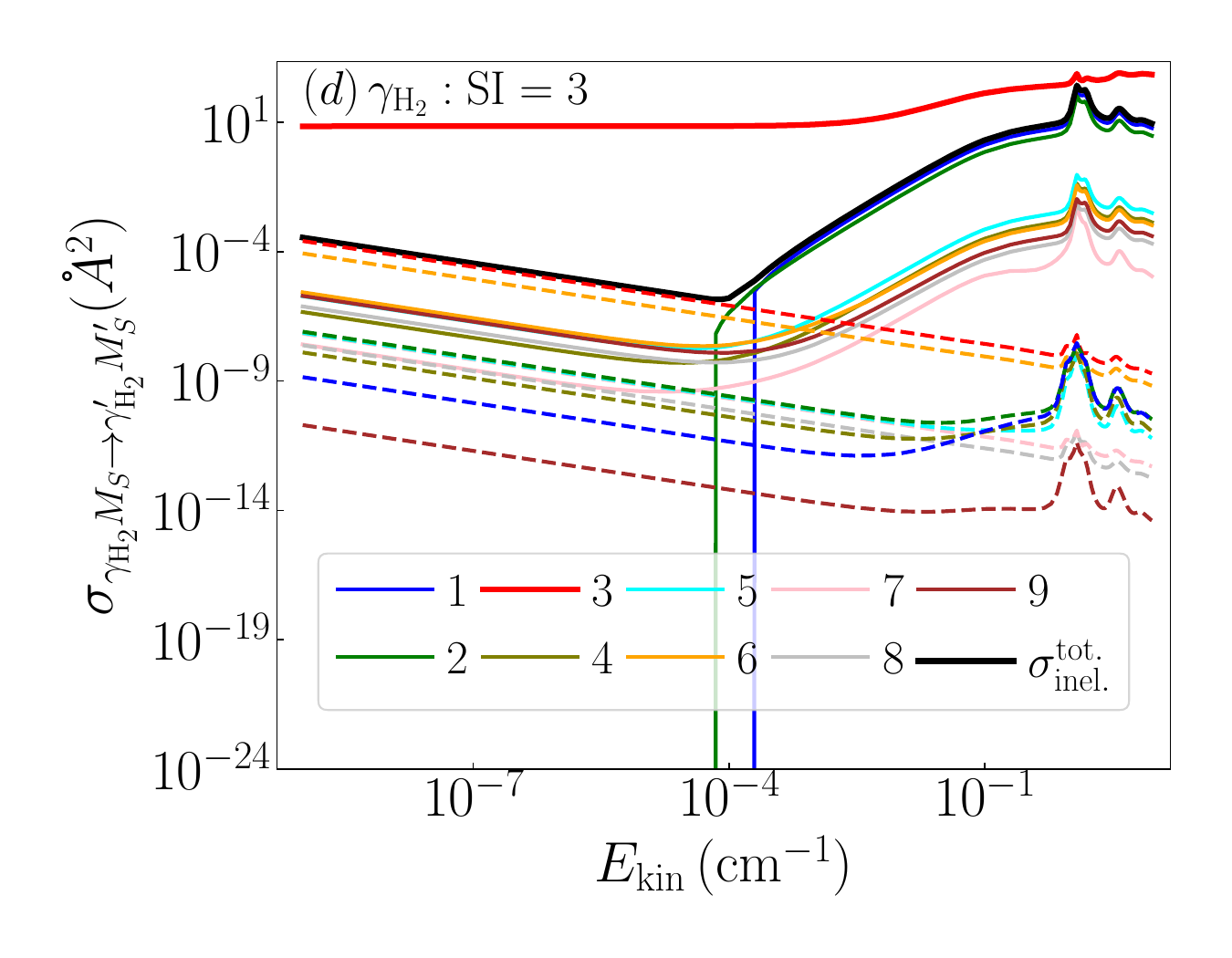}
    \caption{Panel ($a$): Zeeman {sub}levels of the $\nu=0, N=1$ {rovibrational} state {of} H$_{2}$ as a function of magnetic field $B$. Panels ($b$)-($d$): Cross sections for H$_{2}$-Li collisions in an external magnetic field $B=0.3$~T. The {three} panels correspond to different initial levels of H$_{2}$: the $1^{st}$ (panel ($b$)), $2^{nd}$ (panel ($c$)) and the $3^{rd}$ state (panel ($d$)), according to the state labels in panel $(a)$. Solid lines correspond to {collisional} events, in which the electron spin of $^{6}$Li is conserved, and dashed lines correspond to {those}, in which the electron spin {is} flipped ($M_{S} = 1/2 \rightarrow M_{S}' = -1/2$).}
    \label{fig:fig3}
\end{figure}

{{W}hile we {use} the $M_{I}$ and $M_{N}$ quantum numbers to describe {the Zeeman} states at high fields, it is important to acknowledge that at lower fields, the eigenstates undergo significant mixing due to the nuclear spin -- rotation and nuclear spin -- nuclear spin interactions. This situation requires us to use a more general approach for labeling and referencing {the Zeeman} states throughout {the entire range of magnetic fields}. To {this end}, we introduce a "state index" (SI) which uniquely identifies each eigenstate, ranging from 1 to 9, {as shown in Fig.~\ref{fig:fig3}~(a)}. The three trappable states {thus} have SI = 1, 2 and 3.}

We perform quantum scattering calculations at magnetic field strengths {ranging from $10^{-4}$ to 1~T.} As an example, we discuss the {kinetic energy dependence of the} state-to-state cross-sections for H$_{2}$-Li collisions at $B=0.3$~T, which corresponds to a magnetic trap {depth} of approximately {0.8~mK for the SI = 1 state}. We consider collisions of H$_{2}$ and Li in their low-field seeking states (SI = 1, 2 and 3 for H$_{2}$, and $M_{S} = 1/2$, for Li, respectively). Panels ($b$)-($d$) in Fig.~\ref{fig:fig3} {show} the cross-sections for elastic and all inelastic transitions in H$_{2}$-Li {collisions} at $B=0.3$~T. The color of each curve {matches that of the corresponding Zeeman level plotted in Fig.~\ref{fig:fig3}~(a)}. 

The elastic cross-sections {are} consistently {larger than} the total inelastic cross-sections by at least two orders of magnitude, except in the vicinity of {$E_{\mathrm{kin}}= 1.2$~cm$^{-1}$}, where $\gamma$ decreases to $\sim 3$. For the sake of discussion, we distinguish two specific regimes, namely, {the ultralow collision energy regime ($E_{\rm{kin}} < 10^{-5}$~{cm}$^{-1}$), the {low collision} energy regime ($E_{\rm{kin}} > 10^{-2}$~cm$^{-1}$), and an intermediate regime.}

\subsection{Ultralow collision energy regime}

At ultralow collision energies, inelastic cross-sections follow the {$\sim E_{\mathrm{kin}}^{-1/2}$} behavior predicted by Wigner's threshold law.\cite{Wigner_1948} For the low-field-seeking state with the largest internal energy ({SI = 1, panel $(b)$ in Fig.~\ref{fig:fig3}}), two key events contribute to the total inelastic cross-section. The first is the $\Delta M_{I} = \Delta M_{N} = 0, \Delta M_{S} = -1$ transition, i.e., the relaxation of lithium's electron spin, {with H$_{2}$ remaining} in the same Zeeman state. Although this process {clearly} does not lead to the transition of H$_{2}$ to an untrappable state, it involves {the} release of {a} {large} {amount of} energy ($g_{S}\mu_{B}B_{Z} \approx 0.4$~K), which {will induce H$_{2}$ loss}. The second is a nuclear-spin-conserving $\Delta M_{I} = 0$, $\Delta M_{N} = 2$ transition ({$\Delta M_{S} = 0$, see the solid red curve}). While this process leads to the loss of H$_{2}$ population {from} the SI = 1 state, the molecule remains in one of the low-field seeking states after the collision. The process {releases} {$\sim 0.1$~mK} {of energy}. The third most {prominent} contribution, albeit smaller by a factor of 2.5, {is the spin-exchange collision that} involves a simultaneous change in the nuclear spin of H$_{2}$, $\Delta M_{I} = +1$ and relaxation of lithium's electron spin, $\Delta M_{S} = -1$, while conserving the projection of rotational angular momentum, $\Delta M_{N} = 0$, {see the dashed olive curve in Fig.~\ref{fig:fig3}($b$)}. Note that this process is driven directly by the spin-dependent H$_{2}$-Li interaction~(Eq.~\eqref{eq:SDpart-dipole}).

For the second low-field-seeking {initial} state {of H$_{2}$} ({SI = 2, panel $(c)$ in Fig.~\ref{fig:fig3}}), the two most significant contributions to the total inelastic cross-sections stem from lithium's spin-flipping transitions ($\Delta M_{S} = -1$), with either no change of {the} H$_{2}$ quantum numbers ($\Delta M_{I} =  \Delta M_{N} = 0$), or a simultaneous change of $\Delta M_{I} = +1$ (with $\Delta M_{N} = 0$), {see the dashed green and blue lines in Fig.~\ref{fig:fig3}$(c)$, respectively}. While the first transition is essentially elastic for H$_{2}$, it releases {an} energy of $g_{S}\mu_{B}B_{Z} \approx 0.4$~K. The second one is another example of nuclear spin -- electron spin exchange driven by the spin-dependent interaction (Eq.~\eqref{eq:SDpart-dipole}). {In both cases, the released energy will remove the H$_{2}$ molecule from the magnetic trap.} The third most important contribution (4-times smaller than the nuclear spin -- electron spin exchange) {comes} from a nuclear spin-conserving ($\Delta M_{I} = 0$) relaxation to the SI = $3$ state (with $\Delta M_{N} = +1$), with no change {in} lithium's electron spin ($\Delta M_{S} = 0$). Note that the transition to the SI = $1$ low-field-seeking state {of H$_{2}$} is only energetically accessible through a simultaneous spin-flip of lithium, and provides a negligible ($10^{-3}$) contribution to the total inelastic cross-section.

{For} the third low-field-seeking {initial} state {of H$_{2}$} ({SI = 3, panel $(d)$ in Fig.~\ref{fig:fig3}}), two processes {make up} 99\% of the total inelastic cross-section. The first {process} conserves the Zeeman state of H$_{2}$ ($\Delta M_{N} = \Delta M_{I} = 0$), but involves {a} spin flip ($\Delta M_{S} = -1$) {accompanied by an} energy release. The second {process} is a nuclear spin -- electron spin exchange ($\Delta M_S = -1, \Delta M_I = +1$), which conserves $M_{N}$. Note that $M_{I}$-conserving transitions that do not involve a simultaneous spin flip in lithium are not energetically accessible {at $E_{\mathrm{kin}} \lesssim 10^{-4}$~cm$^{-1}$}.

 In all three cases discussed so far, although the {$M_{I}$, $M_{N}$, or $M_{S}$-changing processes} lead to undesired {energy release and loss of H$_{2}$ population from the trap}, the cross-sections for these processes are {over} four orders of magnitude smaller than the elastic cross-section {at collision energies below $10^{-5}$~cm$^{-1}$}. {This suggests excellent prospects for sympathetic cooling of H$_{2}$ in the low-field-seeking states (SI = 1 -- 3) via collisions with spin-polarized Li atoms in a magnetic trap.}

\subsection{{Low} collision energy regime}

The second regime involves kinetic energies {larger than $10^{-2}$~cm$^{-1}$}. Interestingly, for all three {initial} low-field-seeking states {of H$_{2}$}, the cross-sections fall into three distinct categories. The {dominant} contribution (at the level of 99.9\%) to the inelastic cross-section always comes from $M_{I}$-conserving transitions. The second category provides the contribution at the level of $10^{-3} - 10^{-5}$. For the {SI = $1$} low-field-seeking state {(panel $(b)$ in Fig.~\ref{fig:fig3})}, the second category involves two transitions which alter $M_{I}$ by $+1$, and either conserve $M_{N}$ or change $M_{N}$ by $+1$, and one $\Delta M_{I} = +2$, $\Delta M_{N} = 0$ transition. For the SI = 2 low-field-seeking state {(panel $(c)$ in Fig.~\ref{fig:fig3})}, the second category involves three $\Delta M_{I} = +1$ transitions (with $\Delta M_{N} = -1, 0, 1$) and one $\Delta M_{I} = +2$, $\Delta M_{N} = -1$ transition. Finally, the second category for the SI = $3$ trappable state  {(panel $(d)$ in Fig.~\ref{fig:fig3})}, involves all $\Delta M_{I} = +1, +2$ transitions. In all three cases, the third group encompasses all transitions which affect the electronic spin of lithium ($\Delta M_{S} = -1$). It additionally involves the two $M_{S}$-conserving transitions with $\Delta M_{I} = +2$ for the SI = $1$ ({$1 \rightarrow 8$ and $1 \rightarrow 9$}) and SI = $2$ ({$2 \rightarrow 8$ and $2 \rightarrow 9$}) low-field-seeking states and a $\Delta M_{I} = +1$, $\Delta M_{N} = +2$ transition from the SI = $1$ state ({$1 \rightarrow 6$}).

Our results indicate a clear tendency in favor of $M_{I}$- and $M_{S}$-conserving transitions in an external magnetic field. Similar propensity {rules were observed for $M_{S}$} in cold collisions of $^{40}$CaH($X^{2}\Sigma^{+}, v=0, N=1, M_{N} = 1, M_{S} = 1/2$) molecules with $^{4}$He,\cite{Koyu_2022} and for $M_{I}$ in collisions of $^{13}$CO ($X^{1}\Sigma^{+}, v=0, N=1$) with $^{4}$He.\cite{Hermsmeier_2023} The strong suppression of $M_{I}$- and $M_{S}$-changing collisions in the external magnetic field, can be compared to the electron and nuclear spin selection rules in spectroscopy, $\Delta S = 0$, and $\Delta I=0$.\cite{Jacobs_2005}

An intriguing {feature of} the H$_{2}$-Li system {is} the presence of $\Delta M_{I} = 2$ transitions in the second category (or "group-II" {transitions}, as {defined} in Ref.~\cite{Hermsmeier_2023}).  {In the case of transitions from the SI = 1 and 2 states, it is the $\Delta M_{I} = 2$ transition to the SI = 7 state. {Its relative strength can be attributed to a slight {contribution} of the ${|N=1, M_{N}=1\rangle|I=1, M_{I}=-1\rangle}$ {bare} state {to the SI = 7 state. The} mixing of the ${|N=1, M_{N}=1\rangle|I=1, M_{I}=-1\rangle}$ and ${|N=1, M_{N}=-1\rangle|I=1, M_{I}=1\rangle}$ basis states is driven by the nuclear spin -- nuclear spin interactions between the two protons of the H$_{2}$ molecule (Eq.~\eqref{eq:H2_nsns})}.} We performed additional calculations, where we excluded the intramolecular nuclear spin -- nuclear spin interaction from the asymptotic Hamiltonian (Eq.~\eqref{eq:asympt-H2}), and we found that the cross-sections for $\Delta M_{I} = 2$ transitions decreased by four orders of magnitude. Note that this interaction is absent in the $^{13}$CO molecule, studied in Ref.\cite{Hermsmeier_2023} 

\subsection{Magnetic field dependence of the cross-sections}
\begin{figure}[!ht]
    \centering
    \includegraphics[width=\linewidth]{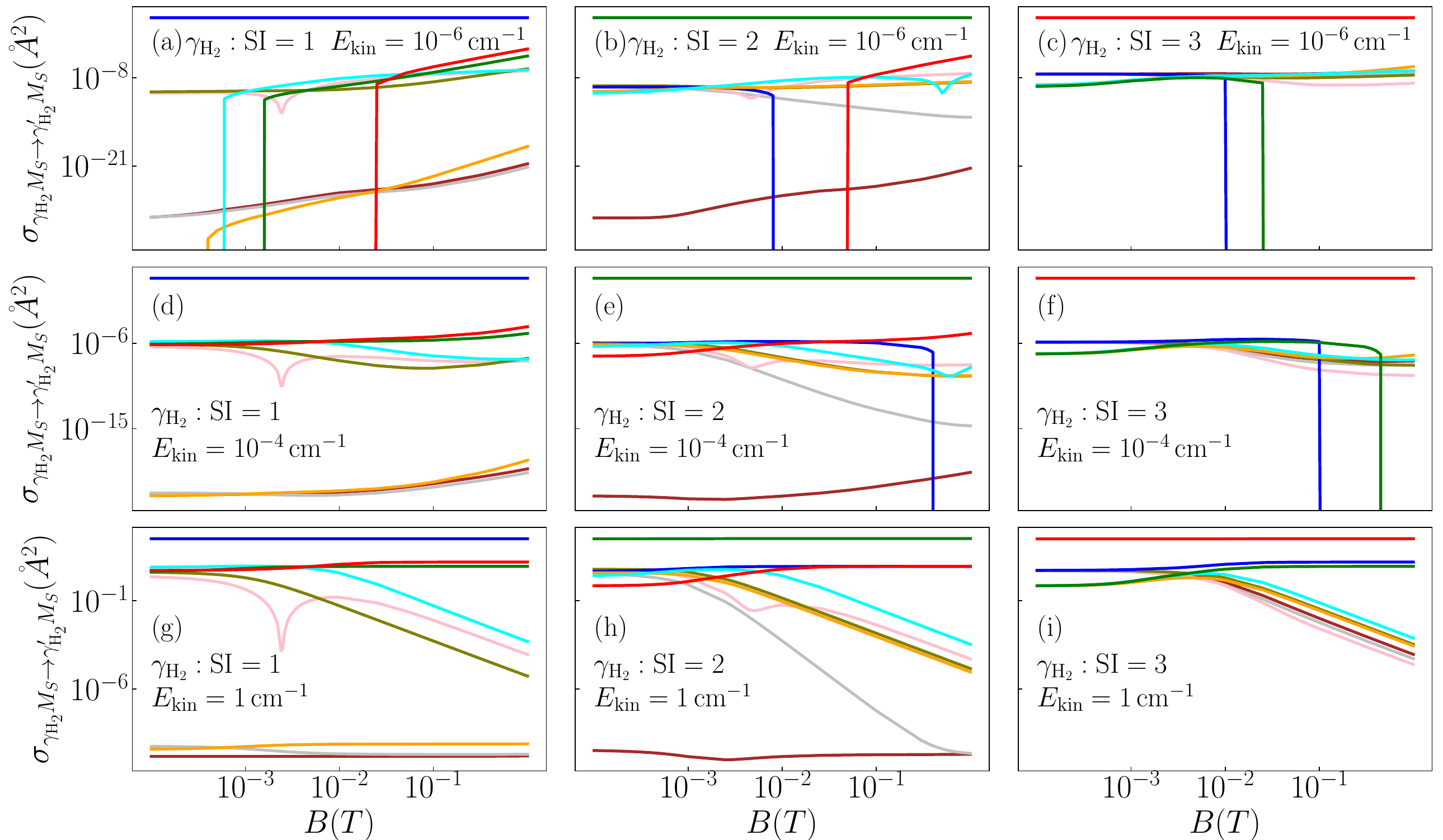}
    \caption{State-to-state cross-sections for {cold} H$_{2}$-Li collisions in the three Zeeman states amenable to magnetic trapping {($\gamma_{\mathrm{H}_{2}}: \mathrm{SI} = 1, 2$ and $3$)}, as a function of the external magnetic field, $B$. The initial and final state of the lithium atom {are} fixed to $M_{S} = 1/2$. {The final Zeeman states in each panel are color-coded according to Fig.~\ref{fig:fig3}.}}
    \label{fig:field_dependence}
\end{figure}
In this subsection, we discuss the {magnetic field} dependence of the state-to-state cross-sections across three distinct collision energy regimes: ultra-low ($10^{-6}$~cm$^{-1}$), intermediate ($10^{-4}$~cm$^{-1}$) and {low} (1~cm$^{-1}$). The discussion builds on {the results in} the previous section, as we focus on collisions of H$_{2}$ in the three {magnetically} trappable states, {SI = 1, 2, and 3, as shown} in Fig.~\ref{fig:field_dependence}. In all panels, the color of each curve corresponds directly to the color of the respective final Zeeman level ({see Fig.~\ref{fig:fig3}}).

{We observe that the e}lastic cross-sections are field-independent and {by far exceed the} inelastic cross-sections. The dependence of the inelastic cross-sections on $B$ {varies with} the kinetic energy and the final Zeeman state. For instance, in the ultra-low energy regime (panels $(a)-(c)$ {in Fig.~\ref{fig:field_dependence}}) we observe a systematic increase of the values of {the cross-sections} with $B$. However, two distinct deviations from this pattern emerge. The first one is related to the observed resonance-like features at 0.0025~T and 0.5~T for the {for the final Zeeman states with} SI = 7 and 5, respectively (see the pink and light blue curves {in {panels ($a$), ($b$), ($d$), ($e$), ($g$), and ($h$) in} Fig.~\ref{fig:field_dependence}}). These will be discussed further in the next paragraph. The other exception is the sharp decline for {excitation} transitions, such as {SI = 2 $\rightarrow$ 1} at 0.008~T ({the blue curve in panel $(b)$}). This {is due to} the closure of inelastic channels due to {increasing} spacing between {the initial and final} Zeeman sublevels with increasing magnetic field. {T}he sharp increase in the cross-sections {(see, for instance, the orange, light blue, and red curves in panel $(a)$)} corresponds to the opening of the {additional inelastic} channels. As the kinetic energy increases (see panels $(d)-(f)$), more channels become energetically accessible, even at {low $B$} field values. {On the other hand,} the fields at which some of the inelastic channels become inaccessible, are shifted towards larger values. Finally, in the {low} energy regime (panels $(g)-(i)$), we can categorize inelastic cross-sections into two main classes. The cross-sections from the first class exhibit {a negligible} field dependence. The cross-sections from the second class are field-independent at low $B$ values, but decrease {monotonically with increasing} $B$ for fields larger than $10^{-2}$~T. We explain this behavior in detail {below}, focusing on the case of the scattering from the SI = 1 state (panel $(g)$).

As discussed in the previous Section, we observe {a} clear propensity for $\Delta M_{I} = 0$ transitions. This {propensity} rule is evident here too, as emphasized by the green and red curves across the considered field ranges: {the} cross-sections for transitions to the SI = 2 and 3 states are notably larger than others. Furthermore, they exhibit {a weak} field dependence. Transitions to the weakly coupled ($\Delta M_{I} = 2$) SI = 8 and 9 states (grey and dark red in panel $(g)$) are orders of magnitude smaller. Interestingly, they are also field-independent, suggesting that the lack of strong coupling makes them less susceptible to the {variations in} $B$. Apart from these two cases, the same observation holds for the transition to the SI = 6 state (orange curve): the relative weakness of this cross-section is related to the {admixture of the $|N=1, M_{N}=0\rangle |I=1, M_{I}=1\rangle$ basis state}. A completely different behavior of the cross-sections as {a} function of $B$ is observed for three other final Zeeman states. Transition to the SI = 4 state (olive line) is one of the most important inelastic processes at low values of $B$. This is because of the {admixture of the $|N=1, M_{N}=-1\rangle |I=1, M_{I}=0\rangle$} state through the nuclear spin-rotation interaction. As the field increases, the energy spacing between the two states decreases, and the mixing becomes less significant. The pronounced magnetic-field dependence of the cross-sections to the Zeeman {eigen}states {composed of strongly mixed bare states $|N M_{N}\rangle |I M_{I}\rangle$} was observed in Ref.~\cite{Hermsmeier_2023} for the $^{13}$C$^{16}$O-He system, and explained in the framework of the Born approximation. For the transitions to the {SI = 5 and 7} states (denoted by the light blue and pink lines), the dynamics are influenced by {the fact that they are composed of three strongly mixed bare states $|N M_{N}\rangle |I M_{I}\rangle$ with $M_{N} + M_{I} = 0$}. This mixing stems from the interplay of the nuclear spin -- rotation and nuclear spin -- nuclear spin interaction (the three states constitute a 3$\times$3 matrix of states with $M_{F} = 0$). The mixing becomes less pronounced as the field increases, although there exists a resonant-like feature at 0.00245~T for the {SI = 1 $\rightarrow$ SI = 7} transition. We note that this feature is independent of the relative kinetic energy of the collision. {The nature of this resonant-like feature and the potential for identifying such resonances in other systems will be explored in a forthcoming {publication}.}

We note similar patterns for inelastic collisions originating from the SI = 2 and 3 states. The majority of significant inelastic processes favor the $\Delta M_{I}=0$ propensity rule, showing only minor variations with increasing fields. For the SI = 2 state, five distinct transitions exhibit a linear decrease with the field. This pattern traces back to the admixture of the {${|N=1, M_{N}=1\rangle|I=1, M_{I}=-1\rangle}$ basis} state, primarily responsible for {the} elevated magnitudes of the inelastic cross-sections at lower fields. As $B$ increases, the mixing {becomes} less pronounced, leading to the decreasing magnitude of the cross-sections at higher $B$ values. Finally, the SI = 3 state is somewhat special: 6 out of 9 cross-sections exhibit a systematic decrease with the field. All of these are "group-II" transitions identified in the previous Section. Their decrease with increasing $B$ can be understood by the decreasing admixture of {the $|N=1, M_{N}=1\rangle|I=1, M_{I}=-1\rangle$ bare state} to the SI = 5 and 7 states (the light blue and pink curves, respectively), the decreasing admixture of the {$|N=1, M_{N}=0\rangle|I=1, M_{I}=-1\rangle$ bare} state {to the SI = $4$ state} (the olive curve), or the increasing energy separation to the rest of the Zeeman states.

\subsection{Elastic-to-inelastic scattering ratio}
In this section, we explore the potential of atomic lithium as {a} sympathetic {coolant for} the H$_{2}$ molecule. While the mismatch {between} Zeeman splittings {of} H$_{2}$ and $^{6}$Li {presents a challenge for the experimental realization of a} two-species trap, here, we focus on estimating the efficiency of the cooling mechanism. Specifically, we determine the elastic-to-inelastic ratio and estimate the optimal lithium density that yields the most effective thermalization during collisions.

\begin{figure}
    \centering
    \includegraphics[width=0.5\linewidth]{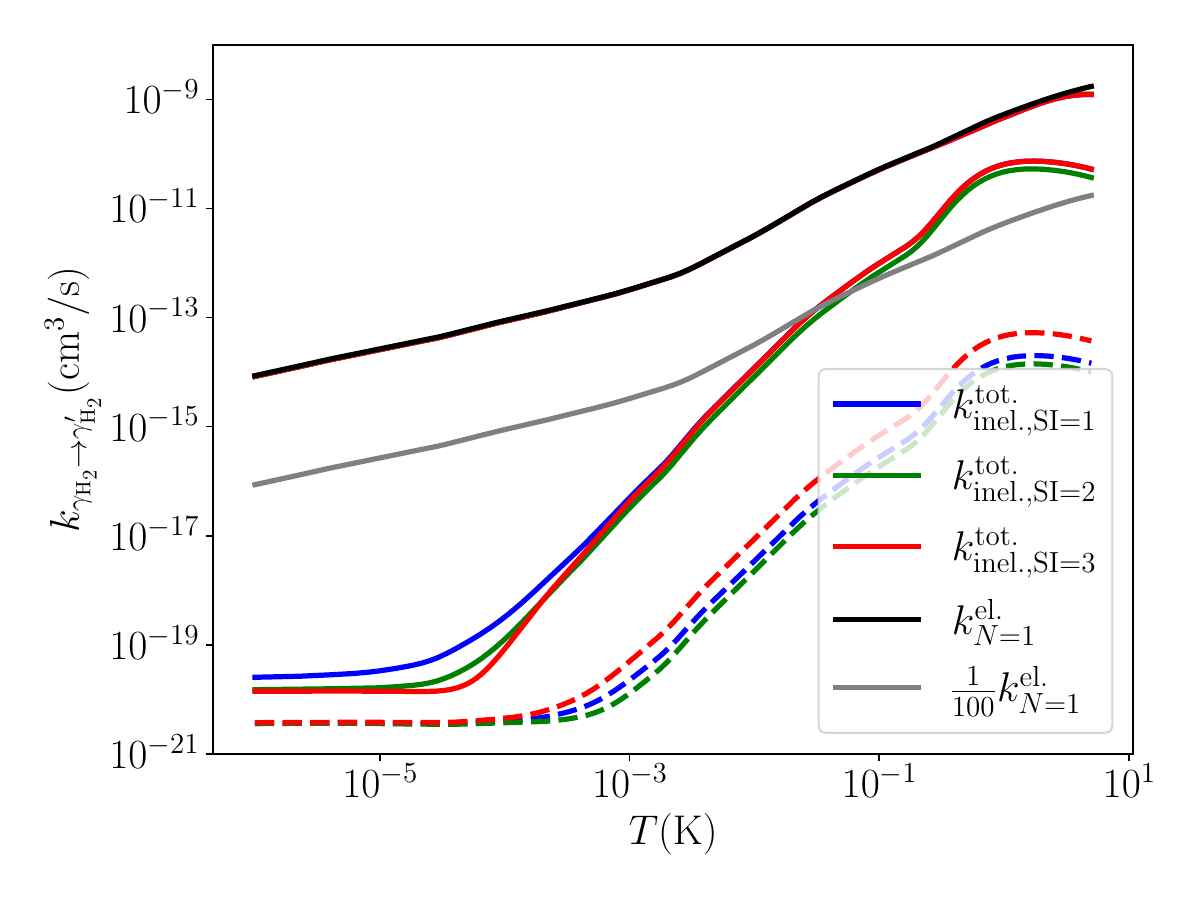}
    \caption{Rate coefficients for elastic and inelastic {transitions in Li -- H$_{2}$} collisions for the three trappable Zeeman sublevels of H$_{2}$ ({$v=0, N=1$}) at 0.3~T. The elastic rates are almost identical to the field- and hyperfine-free elastic rate coefficients for the $v=0, N=1$ state (black solid line). The total inelastic rate coefficients and the inelastic rate coefficients to the untrappable states are presented {as} solid and dashed lines, respectively. The gray solid line presents the elastic rate coefficient {multiplied by $\frac{1}{100}$.}}
    \label{fig:6}
\end{figure}

To this end, we calculate the average state-to-state cross-sections for {Li -- H$_{2}$ collisions for} the three {initial} trappable states (SI = 1, 2 and 3) {of H$_{2}$} (see Eq.~\eqref{eq:simga_av}) and the {corresponding} rate coefficients $k$, {given by}~Eq.~\eqref{eq:k_av}. Note that the initial state of lithium is fixed to the trappable $M_{S} = 1/2$ state.

The rate coefficients are presented in Fig.~\ref{fig:6}. {The e}lastic scattering rates for the three trappable states {of H$_{2}$ are nearly identical to} the corresponding rate coefficients {calculated} without the hyperfine structure and external magnetic field, $k^{\mathrm{el.}}_{N=1}$ (solid black line in Fig.~\ref{fig:6}). The largest difference between $k^{\mathrm{el.}}_{N=1}$ and $k^{\mathrm{el.}}_{\gamma_{\mathrm{H}_{2}}}$ is approximately 3\%. The solid lines in Fig.~\ref{fig:6} correspond to the total inelastic rate coefficients. To check whether the sympathetic cooling of H$_{2}$ {by} $^{6}$Li is feasible, we {plot} in Fig.~\ref{fig:6} the elastic rate coefficient {multiplied by $\frac{1}{100}$ (efficient cooling requires $k^{\mathrm{el.}}/k^{\mathrm{inel.}} \geq 100$).} At first sight, it seems that this condition is fulfilled for temperatures {below} 50~mK. We recall that the proposed trap depth is approximately 0.8~mK for the SI = 1 state of H$_{2}$. However, {the total} inelastic {cross-section is dominated by} transitions to other trappable states (transitions with $\Delta M_{I} = 0$). Thus, the rate of inelastic scattering to states that are \textit{not} amenable {to} magnetic trapping is several orders of magnitude lower (see the dashed lines in Fig.~\ref{fig:6}), and the corresponding elastic-to-inelastic scattering ratio is always larger than $10^{4}$. {Transitions to other trappable states release energy (approximately 10 times lower than the trap depth), which could result in heating, but not trap loss, {being of} minor concern for sympathetic cooling experiments.}

{We also} estimate the rate of thermalizing collisions (in s$^{-1}$) {as}
\begin{equation}
    R = k^{\mathrm{el.}}_{\gamma_{\mathrm{H_{2}}}} n_{0} .
\end{equation}
{T}aking the elastic rate coefficient for the SI = 1 state {of H$_{2}$} at 1~mK ($3.2 \times 10^{-13}$ cm$^{3}$ s$^{-1}$) and the density of $^{6}$Li atoms in the {UV} MOT~\cite{Duarte_2011} operating at 59~$\mu$K ($n_{0} = 2.9 \times 10^{10}$~cm$^{-3}$) we obtain $R \approx 10^{-2}$~s$^{-1}$. For efficient sympathetic cooling, the rate {of} thermalizing collisions should be higher by at least two orders of magnitude. This can be achieved by increasing the density of the lithium MOT, either by using higher magnetic field gradients or larger detunings of the {UV} light. The increased density is then achieved with a tradeoff for an increased temperature of the Li atoms.

\section{Conclusions}
\label{sec:conclusions}
{We performed {a} rigorous {quantum dynamical} analysis of the {effects} of hyperfine and Zeeman interactions on {cold and ultracold} atom -- H$_{2}$ collisions. We investigated cold collisions of molecular hydrogen in the $\nu=0, N=1$ {rovibrational} state with $^{6}$Li {atoms} using CC quantum scattering calculations based on an {accurate} \textit{ab initio} PES. In the field-free case, we found that the three hyperfine levels of the $\nu=0, N=1$ state in H$_{2}$ predominantly undergo elastic collisions. The magnetic dipolar interaction between the electronic spin in lithium and the total nuclear spin of H$_{2}$ {exerts a} pronounced {inference on} ultracold {Li -- H$_{2}$ collisions}, enhancing the $F=0 \rightarrow F'=1$ transitions.}

{{We found that the collisional dynamics of H$_{2}$} in low-field-seeking states in the presence of an external magnetic field {is dominated by} elastic, {rather than} inelastic, collisions. {I}nelastic collisions {tend to} conserve the space-fixed projection of {the} nuclear spin in H$_{2}$. The magnetic dipolar interaction between the nuclear spin of H$_{2}$ and the electronic spin of Li drives the electron spin relaxation {and the nuclear spin -- electron spin exchange, two key inelastic processes} in the ultracold regime.}

{Finally, we discussed the results in the context of the experimental realization of sympathetic cooling of H$_{2}$ {by {ultracold} spin-polarized Li atoms}. Given the {pre}dominance of elastic collisions and the propensity of inelastic scattering to retain H$_{2}$ in {its} low-field-seeking states, {the} elastic-to-inelastic collision ratio for {Li -- H$_{2}$ is favorable for sympathetic cooling ($\gamma > 100$)}. However, to realize efficient sympathetic cooling, the current densities of Li in a MOT must be increased by at least 2 orders of magnitude.}

\section{Acknowledgements}
\label{sec:acknowledgements}
The research is funded by the European Union (ERC-2022-STG , H2TRAP, 101075678). Views and opinions expressed are however those of the author(s) only and do not necessarily reflect those of the European Union or the European Research Council Executive Agency. Neither the European Union nor the granting authority can be held responsible for them. H. J. is supported by the Foundation for Polish Science (FNP). {T.V.T. acknowledges support from the NSF CAREER award No. PHY-2045681.} The research is financed from the budgetary funds on science projected for 2019–2023 as a research project under the “Diamentowy Grant” program.  We gratefully acknowledge Polish high-performance computing infrastructure PLGrid (HPC Centers: ACK Cyfronet AGH, CI TASK) for providing computer facilities and support within computational grant no. PLG/2023/016279. Calculations have been carried out using resources provided by Wroclaw Centre for Networking and Supercomputing (http://wcss.pl), grant no. 546. The research is a part of the program of the National Laboratory FAMO in Toruń, Poland.

\section*{Author Declarations}
\subsection*{Conflict of interest}
The authors have no conflicts to disclose.
    
\subsection*{Author contributions}
\textbf{Hubert J\'{o}\'{z}wiak:} Conceptualization (equal); Investigation (lead); Methodology (equal); Funding acquisition (supporting); Software (equal); Visualization (lead); Writing - Original Draft Preparation (lead).
\textbf{Timur V. Tscherbul:} Conceptualization (equal); Methodology (equal); Funding acquisition (supporting); Resources (supporting); Software (equal); Supervision (supporting); Validation (lead); Writing – review \& editing (equal).
\textbf{Piotr~Wcis{\l}o:} Conceptualization (equal); Funding acquisition (lead); Resources (lead); Supervision (lead); Validation (supporitng); Writing – review \& editing (equal).

\section*{Data availability}
The data that support the findings of this study are available from the corresponding author upon reasonable request.

\newpage
\appendix
\onecolumngrid
\section{Matrix elements in Eq.~\eqref{eq:CC} -- the uncoupled basis}
\label{sec:AppendixA}
{Here, we present a derivation} of the matrix elements that enter the CC equations (Eq.~\eqref{eq:CC}). Following the standard approach~\cite{Krems_2004} we expand the H$_{2}$-Li interaction potential in Legendre polynomials
\begin{equation}
\label{eq:PESexpansion}
    V(\mathbf{R},\mathbf{r}) = \sum_{\lambda = 0}^{\lambda_{\rm{max}}} V_{\lambda}(R,r) P_{\lambda}(\cos\theta) .
\end{equation}
Since H$_{2}$ is a homonuclear molecule, $\lambda$ takes only even values. We truncate the expansion {at} $\lambda_{\rm{max}} = 4$. The interaction potential is diagonal in all spin projections ($M_{I}, M_{S}$, and $M_{I_{\rm{Li}}}$) {with matrix elements}\cite{Krems_2004}
\begin{align}
\begin{split}
    \langle & N M_{N} |\langle I M_{I} | \langle S M_{S}| \langle  l  M_{l} |\hat{V}(\mathbf{R},\mathbf{r}) |N' M_{N}'\rangle |I M_{I}'\rangle |S M_{S}'\rangle | l' M_{l}'\rangle   \\ &=\delta_{M_{S}M_{S}'}\delta_{M_{I}M_{I}'}(-1)^{M_{l}'-M_{N}} \sqrt{[N, N', l, l']} \\
    &\times \sum_{\lambda = 0}^{\lambda_{\rm{max}}} {v_{\lambda,v=0}^{N,N'}(R)} 
    \begin{pmatrix}
        l & \lambda & l' \\
        0 & 0 & 0
    \end{pmatrix}
    \begin{pmatrix}
        l & \lambda & l' \\
       -M_{l} & \Delta M_{l} & M_{l}'
    \end{pmatrix}
    \begin{pmatrix}
        N & \lambda & N' \\
       0 & 0 & 0
    \end{pmatrix}
    \begin{pmatrix}
        N & \lambda & N' \\
       -M_{N} & \Delta M_{N} & M_{N}'
    \end{pmatrix} .
\end{split}
\end{align}
Here, $\Delta M_{x} = M_{x} - M_{x}'$ for all angular momentum projections ($x = l, N, S, I$), and $[x_{1}, x_{2}, ..., x_{N}] = (2x_{1}+1)(2x_{2}+1)...(2x_{N}+1)$. We note that the interaction potential mixes states with different $M_{l}$ and $M_{N}$. At the same time, the interaction conserves the sum $M_{l} + M_{N}$, and, as a result, the projection of the total angular momentum, $M$. {The coefficients $v_{\lambda,v=0}^{N,N'}(R)$ are obtained by taking the matrix elements of the Legendre moments in Eq.~\eqref{eq:PESexpansion}, $V_{\lambda}(R,r)$, between the rovibrational wave functions of the H$_{2}$ molecule in the $v=0$ state}
\begin{equation}
    \label{eq:rovibaverage}
    {
    v_{\lambda,v=0}^{N,N'}(R)  = \int_{0}^{\infty} \mathrm{d}r \chi_{v=0, N}(r)V_{\lambda}(R,r)\chi_{v=0, N'}(r).}
\end{equation}
{Rovibrational wave functions
of H$_{2}$, $\chi_{v, N}$, are obtained by solving the Schrödinger equation for the nuclear motion of H$_{2}$ with the potential energy curve of Schwenke\cite{Schwenke_1988} using the Discrete Variable Representation -- Finite Basis Representation method. Due to a weak dependence of the $v_{\lambda,v=0}^{N,N'}(R)$ terms on $N$, we use $N=N'=1$ in scattering calculations. }

 The magnetic {dipolar interaction between the nuclear spin of H$_{2}$ and the electron spin of Li (see Eq.~\eqref{eq:SDpart-dipole})} is diagonal in $N$ and $M_{N}$
\begin{align}
\begin{split}
    \langle N M_{N} |&\langle I M_{I} | \langle S M_{S}| \langle  l  M_{l} | \hat{V}_{\rm{SD}}(\mathbf{R},\mathbf{r},\hat{\mathbf{I}},\hat{\mathbf{S}}) |N' M_{N}'\rangle |I M_{I}'\rangle |S M_{S}'\rangle | l' M_{l}'\rangle \\ &=  -\delta_{NN'}\delta_{M_{N}M_{N}'} g_{S}\mu_{B}g_{\mathrm{H}}\mu_{N}\Bigl(\frac{\alpha^{2}}{R^{3}} \Bigr) \sqrt{30} (-1)^{-M_{l}+I-M_{\rm{H_{2}}}+S-M_{S}} \sqrt{[l,l']}     
    \begin{pmatrix}
        l & 2 & l' \\
        0 & 0 & 0
    \end{pmatrix} \\ &\times\sqrt{I(I+1)(2I+1)}\sqrt{S(S+1)(2S+1)} 
    \begin{pmatrix}
        1 & 1 & 2 \\
       \Delta M_{S} & \Delta M_{I} & \Delta M_{l}
    \end{pmatrix}\\
    & \times
    \begin{pmatrix}
        I & 1 & I' \\
       -M_{I} & \Delta M_{I} & M_{I}'
    \end{pmatrix}
    \begin{pmatrix}
        S & 1 & S' \\
       -M_{S} & \Delta M_{S} & M_{S}'
    \end{pmatrix}
    \begin{pmatrix}
        l & 2 & l' \\
       -M_{l} & \Delta M_{l} & M_{l}'
    \end{pmatrix} .
\end{split}
\end{align}
{This interaction mixes basis states} with different $M_{S}, M_{I}$ and $M_{l}$, but it conserves the sum $M_{S} + M_{I} + M_{l}$. Thus, the total angular momentum, $M$ is also conserved.

In the next step, we consider the asymptotic Hamiltonian, Eq.~\eqref{eq:asymptHamiltonian}, and we begin with the part of this operator associated with H$_{2}$. The rotational term is diagonal in all quantum numbers
\begin{align}
\begin{split}
    \langle N M_{N} |&\langle I M_{I} | \langle S M_{S}| \langle  l  M_{l} |  \hat{H}_{\rm{rot}} |N' M_{N}'\rangle |I M_{I}'\rangle |S M_{S}'\rangle | l' M_{l}'\rangle   \\
 = &\delta_{NN'}\delta_{M_{N}M_{N}'}\delta_{M_{I}M_{I}'}\delta_{M_{S}M_{S}'}\delta_{M_{l}M_{l}'} \Bigl[B_{e}N(N+1) - D_{v}N^{2}(N+1)^{2}\Bigr] ,
\end{split}
\end{align}
{and the m}atrix elements of the nuclear spin-rotation {interaction} are given as
\begin{align}
    \begin{split}
        -c_{\rm{nsr}} &\langle N M_{N} |\langle I M_{I} | \langle S M_{S}| \langle  l  M_{l} |  \hat{\mathbf{N}}\cdot \hat{\mathbf{I}} |N' M_{N}'\rangle |I M_{I}'\rangle |S M_{S}'\rangle | l' M_{l}'\rangle \\
        &=  -\delta_{NN'}\delta_{M_{S}M_{S}'}\delta_{ll'}\delta_{M_{l}M_{l}'}\Biggl[\delta_{M_{I}M_{I}'}\delta_{M_{N}M_{N}'}c_{\rm{nsr}}M_{N}M_{I}+\Biggr.\\\Biggl.&+\delta_{M_{N}M_{N}'\pm1}\delta_{M_{I}M_{I}'\mp1}\frac{c_{\rm{nsr}}}{2}\Bigl(N(N+1)-M_{N}'(M_{N}'\pm1)\Bigr)^{1/2}\Bigl(I(I+1)-M_{I}'(M_{I}'\mp1)\Bigr)^{1/2} \Biggr].
    \end{split}
\end{align}
The intramolecular {spin -- spin} interaction couples (very weakly) states with different rotational angular momenta
\begin{align}
    \begin{split}
    \label{eq:H2_nsns}
        g_{\rm{H}}^{2}&\mu_{\rm{N}}^{2}\alpha^{2}\langle N M_{N} |\langle I M_{I} | \langle S M_{S}| \langle  l  M_{l} |     \Biggl(\frac{\hat{\mathbf{I}}_{1}\cdot\mathbf{I}_{2}}{r^{3}}-\frac{3(\hat{\mathbf{I}}_{1}\cdot\mathbf{r})(\hat{\mathbf{I}}_{2}\cdot\mathbf{r})}{r^{5}}\Biggr) |N' M_{N}'\rangle |I M_{I}'\rangle |S M_{S}'\rangle | l' M_{l}'\rangle \\ &= \delta_{M_{S}M_{S}'}\delta_{ll'}\delta_{M_{l}M_{l}'}(-1)^{I-M_{I}-M_{N}'}\sqrt{30}c_{\rm{dip}}
        [I]\sqrt{[N,N']} 
        \\
        & \times
        \begin{pmatrix}
            N & 2 & N' \\
            0 & 0 & 0
        \end{pmatrix}
        \begin{pmatrix}
            N & 2 & N' \\
            -M_{N} & \Delta M_{N} & M_{N}'
        \end{pmatrix}
        \begin{pmatrix}
            I & 2 & I \\
            -M_{I} & \Delta M_{I} & M_{I}'
        \end{pmatrix}
        \begin{Bmatrix}
            I_{1} & I_{1} & 1 \\
            I_{2} & I_{2} & 1 \\
            I     & I     & 2 \\
        \end{Bmatrix} \\
        &\times\sqrt{I_{1}(I_{1}+1)(2I_{1}+1)I_{2}(I_{2}+1)(2I_{2}+1)}.
    \end{split}
\end{align}
Here, $\begin{Bmatrix}
            . & . & . \\
            . & . & . \\
            . & . & . \\
        \end{Bmatrix}$ denotes the Wigner 9-j symbol. Both {of the} Zeeman terms in the asymptotic Hamiltonian {of} H$_{2}$ are diagonal in all quantum numbers:
\begin{align}
    \begin{split}
       \langle N M_{N} | &\langle I M_{I} | \langle S M_{S}| \langle  l  M_{l} |     \hat{H}_{\rm{Zeeman}} |N' M_{N}'\rangle |I M_{I}'\rangle |S M_{S}'\rangle | l' M_{l}'\rangle \\ &= -\delta_{NN'}\delta_{M_{N}M_{N}'}\delta_{M_{I}M_{I}'}\delta_{M_{S}M_{S}'}\delta_{M_{l}M_{l}'} \mu_{N}B_{Z}(1-\sigma)\Bigl(g_{r}M_{N} + g_{\rm{H}}M_{I}\Bigr) .
    \end{split}
\end{align}
The same applies to the asymptotic Hamiltonian of Lithium from Eq.~\eqref{eq:asympt-Li}
\begin{align}
    \begin{split}
         \langle N M_{N} | &\langle I M_{I} | \langle S M_{S}|\langle  l  M_{l} |   \hat{H}_{\rm{Li}} |N' M_{N}'\rangle |I M_{I}'\rangle |S M_{S}'\rangle | l' M_{l}'\rangle  \\ &= -\delta_{NN'}\delta_{M_{N}M_{N}'}\delta_{M_{I}M_{I}'}\delta_{M_{S}M_{S}'}\delta_{M_{l}M_{l}'} g_{\rm{S}}\mu_{B}B_{Z} M_{S}.
    \end{split}
\end{align}

\section{Matrix elements in Eq.~\eqref{eq:CC} -- the basis with coupled H$_{2}$ vectors}
\label{sec:appendixB}
Similarly to the uncoupled case, the  H$_{2}$-Li interaction is diagonal in $M_{S}$
\begin{align}
\begin{split}
\label{eq:coupled_ME_pes}
  \langle &(NI)F M_{F}|  \langle S M_{S}|\langle  l  M_{l} | \hat{V}(\mathbf{R},\mathbf{r}) |(N'I)F' M_{F}'\rangle |S M_{S}'\rangle | l' M_{l}'\rangle  \\
  &= \delta_{M_{S}M_{S}'}(-1)^{M_{l}'-M_{F}+I+F+F'} \sqrt{[N, N', l, l', F, F']} \\
    &\times \sum_{\lambda = 0}^{\lambda_{\rm{max}}} {v_{\lambda,v=0}^{N,N'}(R)} 
    \begin{pmatrix}
        l & \lambda & l' \\
        0 & 0 & 0
    \end{pmatrix}
    \begin{pmatrix}
        l & \lambda & l' \\
       -M_{l} & \Delta M_{l} & M_{l}'
    \end{pmatrix}
    \begin{pmatrix}
        N & \lambda & N' \\
       0 & 0 & 0
    \end{pmatrix}\\&\times
    \begin{pmatrix}
        F & \lambda & F' \\
       -M_{F} & \Delta M_{F} & M_{F}'
    \end{pmatrix} 
    \begin{Bmatrix}
        N' & F' & I \\
        F  & N  & \lambda
    \end{Bmatrix} .
\end{split}
\end{align}
The spin-dependent interaction is diagonal in $N$
\begin{align}
\begin{split}
   \langle &(NI)F M_{F}|  \langle S M_{S}|\langle  l  M_{l} | \hat{V}_{\rm{SD}}(\mathbf{R},\mathbf{r},\hat{\mathbf{I}},\hat{\mathbf{S}}) |(N'I)F' M_{F}'\rangle |S M_{S}'\rangle | l' M_{l}'\rangle \\ & = \delta_{NN'} g_{S}\mu_{B}g_{\mathrm{H}}\mu_{N}\Bigl(\frac{\alpha^{2}}{R^{3}} \Bigr) \sqrt{30} (-1)^{M_{l}+I+N+F} \sqrt{[l,l',F,F']}     
    \begin{pmatrix}
        l & 2 & l' \\
        0 & 0 & 0
    \end{pmatrix} \\ &\times\sqrt{I(I+1)(2I+1)}\sqrt{S(S+1)(2S+1)} 
    \begin{pmatrix}
        1 & 1 & 2 \\
       \Delta M_{F} & \Delta M_{S} & \Delta M_{l}
    \end{pmatrix}\\
    & \times
    \begin{pmatrix}
        F & 1 & F' \\
       -M_{F} & \Delta M_{F} & M_{F}'
    \end{pmatrix}
    \begin{pmatrix}
        S & 1 & S' \\
       -M_{S} & \Delta M_{S} & M_{S}'
    \end{pmatrix}
    \begin{pmatrix}
        l & 2 & l' \\
       -M_{l} & \Delta M_{l} & M_{l}'
    \end{pmatrix} 
    \begin{Bmatrix}
        I & F' & N \\
        F & I & 1 
    \end{Bmatrix}.
\end{split}
\end{align}
The rotational part of the Hamiltonian is diagonal in all quantum numbers
\begin{align}
\begin{split}
     \langle &(NI)F M_{F}|  \langle S M_{S}|\langle  l  M_{l} |  \hat{H}_{\rm{rot}} |(N'I)F' M_{F}'\rangle |S M_{S}'\rangle | l' M_{l}'\rangle \\ & = \delta_{NN'}\delta_{FF'}\delta_{M_{F}M_{F}'}\delta_{M_{S}M_{S}'}\delta_{M_{l}M_{l}'} \Bigl(B_{e}N(N+1)-D_{v}N^{2}(N+1)^{2}\Bigr) .
\end{split}
\end{align}
Both hyperfine interactions are diagonal in the total angular momentum of H$_{2}$ and its projection on the space-fixed $Z$-axis. The nuclear spin-rotation interaction is additionally diagonal in all other quantum numbers
\begin{align}
\begin{split}
    -c_{\rm{nsr}} \langle &(NI)F M_{F}|  \langle S M_{S}|\langle  l  M_{l} |  \hat{\mathbf{N}}\cdot \hat{\mathbf{I}} |(N'I)F' M_{F}'\rangle |S M_{S}'\rangle | l' M_{l}'\rangle\\
     &=-\delta_{NN'}\delta_{FF'}\delta_{M_{F}M_{F}'}\delta_{M_{S}M_{S}'}\delta_{M_{l}M_{l}'} \frac{c_{\rm{nsr}}}{2}\Bigl( F(F+1) - I(I+1) - N(N+1) \Bigr) .
\end{split}
\end{align}
The spin -- spin magnetic dipole interaction can, in principle, couple states with different $N$ and $I$. {This} coupling is 11 orders of magnitude smaller than the {spacing} between the $N$ and $N'=N\pm2$ {rotational} states {of} H$_{2}$, {and we neglect it here}. We additionally neglect any \textit{ortho}/\textit{para}-H$_{2}$ coupling. {The m}atrix elements {of} this interaction are
\begin{align}
    \begin{split}
        g_{\rm{H}}^{2}&\mu_{\rm{N}}^{2}\alpha^{2}\langle (NI)F M_{F}|  \langle S M_{S}|\langle  l  M_{l} |  \Biggl(\frac{\hat{\mathbf{I}}_{1}\cdot\hat{\mathbf{I}}_{2}}{r^{3}}-\frac{3(\hat{\mathbf{I}}_{1}\cdot\mathbf{r})(\hat{\mathbf{I}}_{2}\cdot\mathbf{r})}{r^{5}}\Biggr) |(N'I)F' M_{F}'\rangle |S M_{S}'\rangle | l' M_{l}'\rangle \\ &= - \delta_{M_{S}M_{S}'}\delta_{ll'}\delta_{FF'}\delta_{M_{F}M_{F}'}
        (-1)^{N+N'+I+F}\sqrt{30}c_{\rm{dip}}
        [I]\sqrt{[N,N']} 
        \\
        & \times
        \begin{pmatrix}
            N & 2 & N' \\
            0 & 0 & 0
        \end{pmatrix}
        \begin{Bmatrix}
            N & N' & 2 \\
            I & I & F
        \end{Bmatrix}
        \begin{Bmatrix}
            I_{1} & I_{1} & 1 \\
            I_{2} & I_{2} & 1 \\
            I     & I     & 2 \\
        \end{Bmatrix}\sqrt{I_{1}(I_{1}+1)(2I_{1}+1)I_{2}(I_{2}+1)(2I_{2}+1)}.
    \end{split}
\end{align}
The {matrix elements of} Zeeman term of the lithium atom {are} identical {to those given by Eq.~(A8). Finally, the Zeeman Hamiltonian of H$_{2}$} has the following matrix elements:
\begin{align}
    \begin{split}
       \langle&(NI)F M_{F}|  \langle S M_{S}|\langle  l  M_{l} |   \hat{H}_{\rm{Zeeman}} |(N'I)F' M_{F}'\rangle |S M_{S}'\rangle | l' M_{l}'\rangle  \\ &= \delta_{NN'}\delta_{M_{S}M_{S}'}\delta_{M_{l}M_{l}'} \mu_{N}B_{Z}(1-\sigma) (-1)^{F'-m_{F}+I+N'+F'} \\
        &\times 
        \begin{pmatrix}
            F       & 1 & F'     \\
            -M_{F}  & 0 & M_{F}'
        \end{pmatrix}
        \sqrt{[F,F']}\sqrt{I(I+1)(2I+1)}
        \Bigl(
        \delta_{NN'} g_{r} 
        \begin{Bmatrix}
        F  & F' & 1 \\
        I  & I  & N 
        \end{Bmatrix}
        + g_{\rm{H}}
        \begin{Bmatrix}
        F  & F' & 1 \\
        N' & N  & I 
        \end{Bmatrix}
        \Bigr).
    \end{split}
\end{align}

\bibliography{bibliography}

\end{document}